\documentclass[a4paper,11pt]{article}
\RequirePackage{snapshot}  

\usepackage{jheppub} 

\usepackage{bm,amsmath,amssymb,slashed,graphicx,%
            enumerate,alltt,xspace,multirow,xcolor,mathrsfs}
\usepackage{fancyvrb}
\usepackage{booktabs,tabularx}
\usepackage{graphicx}
\usepackage{subcaption}
\usepackage{xspace}
\usepackage[utf8]{inputenc} 
\usepackage[compat=1.0.0]{tikz-feynman}
\usepackage{scalerel}
\usepackage{layouts}

\usepackage{lineno}
\usepackage{printlen}
\usepackage{wasysym}
\usepackage{grffile}

\makeatletter
\g@addto@macro\bfseries{\boldmath}
\makeatother

\definecolor{labelkey}{rgb}{0,0.5,0.0}

\usepackage{listings}
\definecolor{darkgreen}{rgb}{0,0.4,0}
\definecolor{lightblue}{rgb}{0.0,0.5,1.0}
\definecolor{grey}{rgb}{0.5,0.5,0.5}

\allowdisplaybreaks

\definecolor{semiblue}{rgb}{0.3,0.3,0.8}
\newcommand{\logbook}[2]{}

\newcommand{\betaps}{{\beta_\text{\textsc{ps}}}}

\newcommand{\muR}{\mu_\text{\textsc{r}}}
\newcommand{\muF}{\mu_\text{\textsc{f}}}
\newcommand{\xR}{x_\text{\textsc{r}}}
\newcommand{\xF}{x_\text{\textsc{f}}}

\newcommand{\GeV}{\;\mathrm{GeV}} \newcommand{\TeV}{\;\mathrm{TeV}}

\newcommand{\as}{\alpha_s}
\newcommand{\pdf}{\text{\textsc{pdf}}}
\newcommand{\aspdf}{\alpha_{s,\pdf}}

 \newcommand{\nc}{N_\text{\textsc{c}}}

\newcommand{\itilde}{{\tilde \imath}} \newcommand{\jtilde}{{\tilde \jmath}}

\newcommand{\showercut}{p_{t,\text{cut}}}
\newcommand{\ifinal}{{\hat \imath}}
\newcommand{\iorigin}{i}
\newcommand{\xfinal}{{\hat x}}
\newcommand{\xorigin}{x}

\tikzstyle{block} = [rectangle, minimum width=1.0cm, minimum height=0.75cm,
thin, draw=black] \tikzstyle{blob} = [circle, minimum width=0.5cm, thin,
draw=black] \tikzset{blackarrow/.style={-stealth, semithick, draw=black}}
\tikzset{connection/.style={inner sep=0,outer sep=0}}

\newcolumntype{C}{>{\centering\arraybackslash}X}

\title{PanScales showers for hadron collisions: all-order validation }

\preprint{OUTP-22-10P}

\newcommand{\OXaff}{Rudolf Peierls Centre for Theoretical Physics, Clarendon
  Laboratory, Parks Road, University of Oxford, Oxford OX1 3PU, UK}
\newcommand{\ASCaff}{All Souls College, Oxford OX1 4AL, UK}

\author[a]{Melissa van Beekveld,}%
\author[a]{Silvia Ferrario Ravasio,}%
\author[b]{Keith Hamilton,}%
\author[a,c]{Gavin P.~Salam,}%
\author[d]{Alba~Soto-Ontoso,}%
\author[d]{Gregory Soyez,}%
\author[b]{Rob Verheyen}%

\affiliation[a]{\OXaff}
\affiliation[b]{Department of Physics and Astronomy, University College London,
London, WC1E 6BT, UK}
\affiliation[c]{\ASCaff}
\affiliation[d]{Universit\'e Paris-Saclay, CNRS, CEA, Institut de physique
th\'eorique, 91191, Gif-sur-Yvette, France}

\date{Received: date / Accepted: \today}

\abstract{
  We carry out extensive tests of the next-to-leading logarithmic (NLL) accuracy of the
  PanScales parton showers, as introduced recently for colour-singlet
  production in hadron collisions. 
  The tests include comparisons to (semi-)analytic NLL calculations of
  a wide range of hadron-collider observables:
  the colour-singlet boson transverse momentum distribution;
  global and non-global hadronic energy flow variables related to jet
  vetoes and analogues of jettiness distributions;
  (sub)jet multiplicities;
  and observables sensitive to the DGLAP evolution of the incoming momentum
  fractions.
  In the tests, we also include an implementation of a standard
  transverse-momentum ordered dipole shower, to establish the size of
  missing NLL effects in such showers, which, depending on the
  observable, can reach $100\%$.
  This paper, together with~\cite{vanBeekveld:2022zhl}, constitutes the first
  step towards process-independent NLL-accurate parton showers for hadronic
  collisions.
} 


\begin{document}

\maketitle

\section{Introduction}
\label{sec:intro}

Parton-shower simulations lie at the core of the majority of experimental and
phenomenological studies in collider physics,
accounting for the physics of parton branching
across several orders of magnitude in momentum scale, independently of
any specific observable.
As such, one of the key questions, for both existing and new parton
showers, is to understand and demonstrate their accuracy as compared
to the standard QCD tool for multi-scale problems, namely logarithmic
resummation. 
In a companion paper~\cite{vanBeekveld:2022zhl}, we recently
formulated new classes of initial-state parton showers (PanGlobal and
PanLocal) specifically designed to achieve next-to-leading logarithmic
(NLL) accuracy in the context of hadron-hadron collisions.
That paper included a number of tests of the kinematic recoil
properties of the shower in the presence of two or three emissions,
and validation against exact fixed-order matrix elements for spin and
colour degrees of freedom.
Those tests provided strong evidence that the new showers resolve key problems
that are found in a
standard~\cite{Sjostrand:2006za,Giele:2007di,Schumann:2007mg,Platzer:2009jq,Hoche:2015sya,Cabouat:2017rzi}
transverse-momentum ordered dipole approach, problems similar to those observed some time
ago in final-state showers~\cite{Hamilton:2020rcu} and related to long-standing
discussions about the treatment of initial-state recoil~\cite{Nagy:2009vg,
Platzer:2009jq, Hoche:2015sya, Cabouat:2017rzi}.

In this paper we present a number of all-order logarithmic tests in
the context of colour-singlet production in proton--proton collisions.
We test the new PanScales showers and, for the purpose of comparison,
our implementation of a standard dipole shower, which we refer to as
Dipole-$k_t$.
These are the first all-order logarithmic tests to be carried out for
initial-state showers, extending the developing body of recent work
for final-state showers~\cite{Dasgupta:2018nvj,Dasgupta:2020fwr,Hamilton:2020rcu,Karlberg:2021kwr,Hamilton:2021dyz}.
The tests serve two purposes.
Firstly, they provide verification of the NLL accuracy for the PanScales
showers across a wide range of observables, for an arbitrary number of
emissions and taking into account all-order evolution of the strong
coupling and the parton distribution functions (PDFs).
Secondly, for showers that are not NLL accurate for a specific
observable, they enable us to quantify the size of the deviation from
the NLL result.
While we will not go so far as to examine detailed phenomenological
consequences in this paper,\footnote{To do so would require matching
  with fixed-order and possibly an interface to hadronisation,
  neither of which are currently available within the PanScales approach
  for hadron-collider processes.} for each of the observables that we
consider, we will comment on how it relates to widely discussed
phenomenological questions.

We start our discussion with a brief review of the showers that we
consider (Section~\ref{sec:basic-formulation}) and then turn to a
number of observables.
One critical new test relative to the final-state case is the
verification of the accuracy of PDF evolution
(Section~\ref{sec:pdf-origin}), and we comment briefly also on a
practical observable that could be used for related measurements in
data.
We then consider a variety of global event quantities with distinct
resummation structures.
These include the
jet-veto acceptance probability and observables related to
0-jettiness~\cite{Stewart:2010tn} (Section~\ref{sec:global-obs}), for
which the results are qualitatively similar (and in some cases
quantitatively identical) to corresponding final-state tests.
We then turn our attention to a particularly important global observable, the
colour-singlet transverse momentum distribution
(Section~\ref{sec:transverse-mom}), for which we test not just the Sudakov
region, but also the characteristic
power-suppressed region identified long ago by Parisi and
Petronzio~\cite{Parisi:133268}.
Then follow tests of energy flows in limited angular regions
(Section~\ref{sec:rapidity-slice}), which play a role in many collider
contexts, and a study of another basic observable, the average
particle multiplicity (Section~\ref{sec:multiplicity}).
We conclude with some exploratory phenomenological studies of the
impact of our NLL showers on the $Z$-boson transverse momentum
distribution and on the azimuthal correlations of jets (Section~\ref{sec:exploratory-pheno}).

\section{Brief overview of the showers and the testing approach}
\label{sec:basic-formulation} 

Throughout we consider the production of a colourless boson in proton--proton
collisions, either $\bar{q}(\tilde{p}_a) q(\tilde{p}_b) \to Z$ or
$g(\tilde{p}_a)g(\tilde{p}_b)\to H$, at a proton--proton centre-of-mass energy $\sqrt s$ and
with Born invariant mass squared $m_X^2 = (\tilde{p}_a + \tilde{p}_b)^2$. The 4-momentum of the
colour-singlet (hard system) is defined as 
\begin{equation}
	Q^{\mu} =  m_X (\cosh y_X, 0,0, \sinh y_X)\,,
\end{equation}
where $X = Z, H$, and $y_X$ denotes the rapidity of the hard system. 
All partons are
considered to be massless. We will compare the all-order behaviour of
a standard dipole shower, which we refer to as Dipole-$k_t$, and the PanScales showers
introduced in Ref.~\cite{vanBeekveld:2022zhl} suitable for hadron-hadron
collisions. Here we give a brief summary of these showers, and full details can be
found in Ref.~\cite{vanBeekveld:2022zhl}.

For all showers, the momentum of a newly emitted parton $k$ is
decomposed as
\begin{equation}
  \label{eq:pk}
  p_k = a_k \tilde{p}_i + b_k \tilde{p}_j + k_\perp,
\end{equation}
where $\tilde{p}_{i,j}$ are the pre-branching momenta of the dipole
constituents. By convention, $i$ labels the emitter and $j$ the spectator. The
vector $k_\perp$ is space-like, orthogonal to $\tilde{p}_{i,j}$ and satisfies
$k_\perp^2= - 2 a_k b_k \tilde{p}_i \cdot \tilde{p}_j$. The coefficients $a_k$
and $b_k$ are related to a shower-specific ordering variable $v$ and an
auxiliary rapidity-like variable $\bar{\eta}$.

\paragraph{Dipole-$k_t$ showers:} 
our Dipole-$k_t$ class of showers follows in the long line of dipole
showers inspired by
Refs.~\cite{Gustafson:1987rq,Catani:1996vz,Catani:2002hc}.
It shares substantial similarities with the dipole showers
available in all the major Monte Carlo event generators,
e.g.~Pythia~\cite{Cabouat:2017rzi},\footnote{Specifically the shower
  with local recoil for initial-final dipoles, which is not its
  default.}
Sherpa~\cite{Schumann:2007mg} and
Herwig~\cite{Platzer:2009jq}.
In the soft-collinear limit, the ordering variable $v$ corresponds to
the transverse momentum of the emission $|k_\perp|$.

The recoil scheme for emissions from final-final (FF) or final-initial (FI)
dipoles is fully dipole-local, i.e.
\begin{subequations}
  \label{eq:dipole-local-scheme}
  \begin{align}
    p_i &= a_i \tilde{p}_i + b_i\tilde{p}_j - k_\perp\,,\\
    p_j &=  b_j \tilde{p}_j\,,
  \end{align}
\end{subequations}
where
the coefficients $a_i, b_i$ and $b_j$ can be related to $a_k$ and
$b_k$  using $p_{i,j}^2 =0$ and $p_i\pm p_j+p_k = 
	\tilde{p}_i\pm \tilde{p}_j$, taking the $+$ sign if the
recoiler $j$ is in the final state, $-$ otherwise. 
For emissions from initial-initial (II) dipoles, the recoil is instead distributed globally, i.e.
  \begin{align}
    p_i = a_i \tilde{p}_i\,, \qquad 
    p_j = \tilde{p}_j\,,
  \end{align}
followed by an event-wide boost (excluding the last emitted parton) that
restores momentum conservation.

In the case of emissions from initial-final (IF) dipoles, where the
initial-state parton is identified with the emitter, we consider two recoil
schemes: one local and one global (see e.g.\
Refs.~\cite{Platzer:2009jq,Hoche:2015sya}), reflecting the variety of schemes
implemented in public parton shower codes.
In the fully {\it local} scheme, the transverse recoil is assigned to the final-state
(spectator) parton, exactly like in the FI dipole.
This implies that only II
dipoles can impart transverse momentum recoil to the hard colour-singlet system.
It is well-known that this leads to wrong predictions for the $Z$
transverse momentum distribution at the NLL-level
\cite{Nagy:2009vg, Platzer:2009jq, Hoche:2015sya, Cabouat:2017rzi},
but it remains widely used, hence it will be of interest to
quantify its deviation from the NLL expectation.
%
%
In the {\it global} scheme, the recoil is distributed according to
Eq.~\eqref{eq:dipole-local-scheme} (but with a positive sign for
$k_\perp$).
Next, all the particles in the event are
boosted to realign $p_i$ with the beam axis. This effectively implies that the
transverse recoil is redistributed across the event. 

For any dipole type where the assignment of transverse recoil depends on which
end of the dipole is the emitter, the choice of emitter is based on the end of
the dipole that is closer in angle to the radiation in the dipole centre-of-mass
frame (with a smooth transition between the two regions).
This means that the rapidity-like auxiliary generation variable
$\bar{\eta}$ coincides with the rapidity measured in the
emitting-dipole frame.
%
%

\paragraph{PanScales showers:}
%
in the PanScales showers, we use a class of evolution variables $v$
that is parametrised in terms of a quantity $\betaps$, which determines
the relation between $v$, transverse momentum $\kappa_\perp$ and
rapidity $\bar{\eta}_Q$.
Specifically, we define
\begin{align}
	\label{eq:evolutionv}
	v = \frac{\kappa_\perp}{\rho}{\rm e}^{-\betaps
  |\bar{\eta}_Q|}\,,
  \qquad
  \text{with } \rho = \left(\frac{\tilde{s}_{i} \tilde{s}_{j}}{Q^2 
	\tilde{s}_{ij}}\right)^{\frac{\betaps}{2}},
\end{align}
where $\tilde{s}_{i,j} = 2\tilde{p}_{i,j} \cdot Q$, with
$\tilde{s}_{ij} = 2\tilde{p}_i\cdot \tilde{p}_j$ the dipole mass
squared.
The precise relation between $\kappa_\perp$ and $|\bar{\eta}_Q|$ on
one hand, and the $k_\perp$, $a_k$ and $b_k$ of Eq.~(\ref{eq:pk}) on
the other, depends on the shower, as discussed in
Ref.~\cite{vanBeekveld:2022zhl}.

%
%

The PanScales showers come in two variants, PanLocal and PanGlobal.
PanLocal always employs dipole-local recoil, resembling the global option 
of Dipole-$k_t$. This means
\begin{subequations}
  \begin{align}
    p_i &= a_i \tilde{p}_i + b_i\tilde{p}_j \pm f k_\perp,\\
    p_j &= a_j \tilde{p}_i + b_j\tilde{p}_j \pm (1-f) k_\perp,
  \end{align}
\end{subequations}
where $f=1$ for the PanLocal dipole variant (i.e.\ the emitter takes
the entire transverse recoil of the emitted parton), and $f=\frac{e^{2\bar{\eta}}}{e^{2\bar{\eta}}+1}$ 
in the PanLocal antenna variant (the transverse recoil is shared between the
emitter and the spectator). 
The sign $\pm$ in front of $k_\perp$
depends on whether
the parton is in the initial-state ($+$) or in the final state ($-$).

For a given dipole,
the choice of effective emitter is based on the sign of $\bar{\eta}_Q$
(except in a transition region around $\bar{\eta}_Q=0$), i.e.\
taking the dipole end that is closer in the event frame rather than the
emitting-dipole frame.
All the coefficients in the kinematic map are then fixed
%
by imposing local momentum conservation and that the post-splitting
partons be on shell.
When, following the mapping, an initial-state parton is misaligned
with the beam axis, a Lorentz transform is applied to the whole event
so as to realign it, with the constraint that the hard-system rapidity
is preserved.


In the PanGlobal shower, for all dipole types, only the longitudinal components are conserved locally
\begin{subequations}
  \begin{align}
    p_i = (1\pm a_k) \tilde{p}_i, \\
    p_j = (1\pm b_k) \tilde{p}_i, 
  \end{align}
\end{subequations}
while the transverse recoil is assigned directly to the colour singlet
system.
Further rescalings are then applied to the two initial-state momenta
so as ensure that the hard-system mass and rapidity are preserved.
%

The fixed-order considerations of Ref.~\cite{vanBeekveld:2022zhl} lead us to
expect that PanLocal dipole/antenna with $0 < \betaps < 1$ and PanGlobal with
$0 \leq \betaps < 1$ are NLL accurate.
In this article, we consider the PanLocal dipole and antenna showers
with $\betaps = 0.5$, and the PanGlobal shower with $\betaps = 0$ and
$0.5$.
%

As in earlier PanScales work, we provide an all-order validation of the
logarithmic accuracy of our showers by comparing their predictions to known
resummations.
Considering the logarithm $L$ of some observable, we take one of two limits
(depending on the observable's resummation properties):
$\as L$ fixed with $\as$ approaching $0$, used for checking LL and NLL
accuracy;
or $\as L^2$ fixed with $\as$ approaching $0$, used for checking
double-logarithmic (DL) and next-to-double logarithmic (NDL)
accuracy.
We account for subleading-colour effects, using the NODS method of
Ref.~\cite{Hamilton:2020rcu}, which nests full-colour energy-ordered
double-soft matrix-element corrections.
This is expected to result in full-colour NLL accuracy for all
observables except non-global ones, which have only leading-colour
accuracy for the single-logarithmic (NLL) terms.
Spin correlations for initial-state radiation are included in the
PanScales code using our adaptation and extension of the
Collins-Knowles
algorithm~\cite{Collins:1987cp,Knowles:1987cu,Knowles:1988vs,Knowles:1988hu,Karlberg:2021kwr,
  Hamilton:2021dyz}, as discussed and studied in
Ref.~\cite{vanBeekveld:2022zhl}.
The observables that we examine in this paper are insensitive to spin
correlations at our target NLL/NDL accuracy and our runs are performed
without them.\footnote{
  Spin correlations involve an $\mathcal{O}(1)$ speed cost (the cost
  depends on event multiplicities) and it was beneficial to trade
  that cost for extra statistics. }

When we show results, our errors bands correspond to one standard
deviation ($\sigma$), and showers are considered to pass a given test if
the deviation from the expected accuracy (NLL or NDL) is
$<2\sigma$.\footnote{An astute reader may complain that for $5\%$ of
  tests we therefore expect failure, insofar as the error is dominated
  by statistical effects (which is not always the case).
  When we see a failure that is borderline larger than $2\sigma$, we
  generally generate additional statistics until a clear conclusion
  can be drawn.  }

For most of our observables, a novel feature relative to earlier PanScales work
is the impact of parton distribution functions on NLL and NDL terms. The
handling of the $\as \to 0$ limits in the evolution of the PDFs is the subject
of Appendix~\ref{sec:choice-pdfs}. 
Besides this, the numerical techniques that we use are largely the
same as in previous PanScales
work~\cite{Dasgupta:2020fwr,Hamilton:2020rcu,Karlberg:2021kwr,Hamilton:2021dyz}.
In general we will express our results in terms of quantities such as
$\lambda = \as L$ or $\xi = \as L^2$.
A translation to physical momentum scales is given in Table~1 of
Ref.~\cite{Hamilton:2020rcu}. 

\section{Single-logarithmic comparisons with DGLAP evolution}
\label{sec:pdf-origin}

The first test we perform on our showers is to establish whether they
correctly reproduce DGLAP evolution.
We assume an initial $\bar{d}d \rightarrow Z$ or $gg\rightarrow H$
event, at fixed initial rapidity $y_X$, and perform parton showering down
to an effective transverse momentum cutoff scale, $p_{t,\text{cut}}$.
Focusing on the $Z$ case, let $\ifinal$ be the flavour of one of the two
partons entering the hard scattering process
and $\xfinal$ its momentum fraction,
\begin{equation}
\xfinal_\pm =\frac{m_Z}{ \sqrt{s}} \exp(\pm y_Z)\,,
\end{equation}
where the sign matches the sign of the $z$-momentum of the incoming
parton.
After parton showering, the incoming parton has a flavour $\iorigin$
and a momentum fraction $\xorigin$.
This is the parton effectively extracted from the proton at a
factorisation scale of the order of $p_{t,\text{cut}}$.
Our tests here determine whether, for a given $\ifinal$ and $\xfinal$
the distribution over $\iorigin$, $\xorigin$ matches that expected
from DGLAP evolution.

The distribution over $\iorigin$, $\xorigin$ is dominated by single
logarithmic terms, $\as^n L^n$, where $L = \ln \showercut / m_Z $.
To understand how to determine the expectation, let us introduce
$D_{ij}(z,\as L)$, a single-logarithmic DGLAP evolution operator such
that the PDFs satisfy
\begin{equation}
  f_\ifinal(\xfinal, m_Z^2) =
  \sum_j \int_\xfinal^1 \frac{{\rm d}z}{z} z D_{\ifinal j}(z,\as L)
  f_j\left(\frac{\xfinal}{z}, \showercut^2\right),
\end{equation}
where $f_i(x, \mu^2)$ is the density of partons of flavour $i$, carrying
momentum fraction $x$ at a factorisation scale $\mu$.\footnote{%
  Given an initial condition $D_{ij}(x,0) = \delta_{ij} \delta(1-x)$,
  the evolution operator satisfies the differential equation
  \begin{equation}
    \partial_{\lambda} D_{ij}(x,\lambda)
    =
    \sum_k \frac{1}{\pi} \int_x^1 \frac{{\rm d}z}{z} P_{ik}(z)  D_{kj}(x/z,\lambda)\,.
  \end{equation}
}
The DGLAP expectation for the distribution over the flavour
$\iorigin$, and momentum fraction $\xorigin$ at the shower cutoff
scale is given by
\begin{equation}
  \label{eq:DGLAP-origin}
  \frac{1}{\sigma} \frac{{\rm d}\sigma_{\iorigin}}{{\rm d}\xorigin}
  =
  \frac{1}{f_{\ifinal}(\xfinal, m_Z^2)}
  \int_{\xfinal}^1 \frac{{\rm d}z}{z} D_{\ifinal\iorigin}(z,\as L)
     f_{\iorigin}\left(\frac{\xfinal}{z}, \showercut^2\right)
    \delta\left(
      \frac{\xfinal}{z} - \xorigin\right).
\end{equation}
It is interesting to ask how Eq.~(\ref{eq:DGLAP-origin}) relates to a
physical observable that could be measured at colliders.
Leaving aside the question of flavour, one could imagine clustering an
event with some inclusive jet algorithm, with a jet transverse
momentum threshold $p_{t,\min}$, playing the role of
$p_{t,\text{cut}}$, and then determining the distribution of $x_\pm$
defined as
\begin{equation}
  \label{eq:x-origin-MC}
  \xorigin_\pm = 
  \sum_{i \in X,\,\text{jets}} \frac{E_i \pm p_{z,i}}{E_p \pm p_{z,p}}
  \,.
\end{equation}
Here $X$ is the hard system (for example the Drell-Yan pair), $E_p$
and $p_{z,p}$ are the energy and $z$-momentum of the incoming proton,
and the choice of sign depends on whether one is considering the
proton (and incoming parton direction) with positive or negative
$z$-momentum.
Phenomenologically the distribution of $x_\pm$ is very sensitive to the
pattern of forward-jet radiation.
It is infrared and collinear safe and thus calculable within
perturbation theory.
At single-logarithmic accuracy, the distribution of $x_\pm$ coincides with
Eq.~(\ref{eq:DGLAP-origin}) so long as one chooses
$L = \ln p_{t,\min}/m_Z$ and sums over flavours $\iorigin$ and
$\ifinal$ (the latter with a suitable hard-cross section weight).
One could extend the measurement of $\xorigin_\pm$ to be differential
in the relative azimuthal angles of multiple initial-state hard jets,
which would introduce sensitivity to spin correlations.

\begin{figure}[t]
	\centering
	\begin{subfigure}{\textwidth}
		\includegraphics[page=3, width=\textwidth]{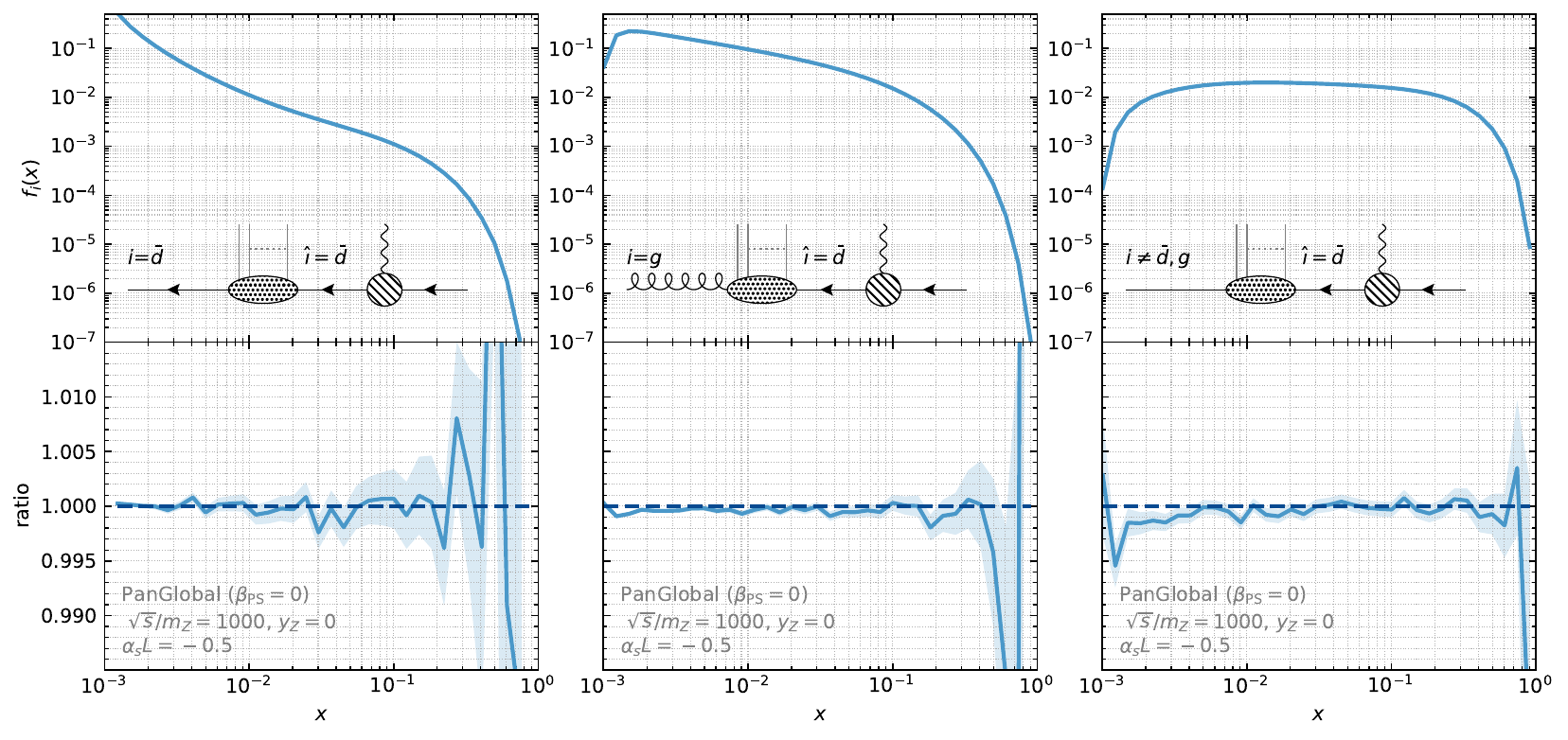}
	\end{subfigure}
	
	\begin{subfigure}{\textwidth}
		\includegraphics[page=5, width=\textwidth]{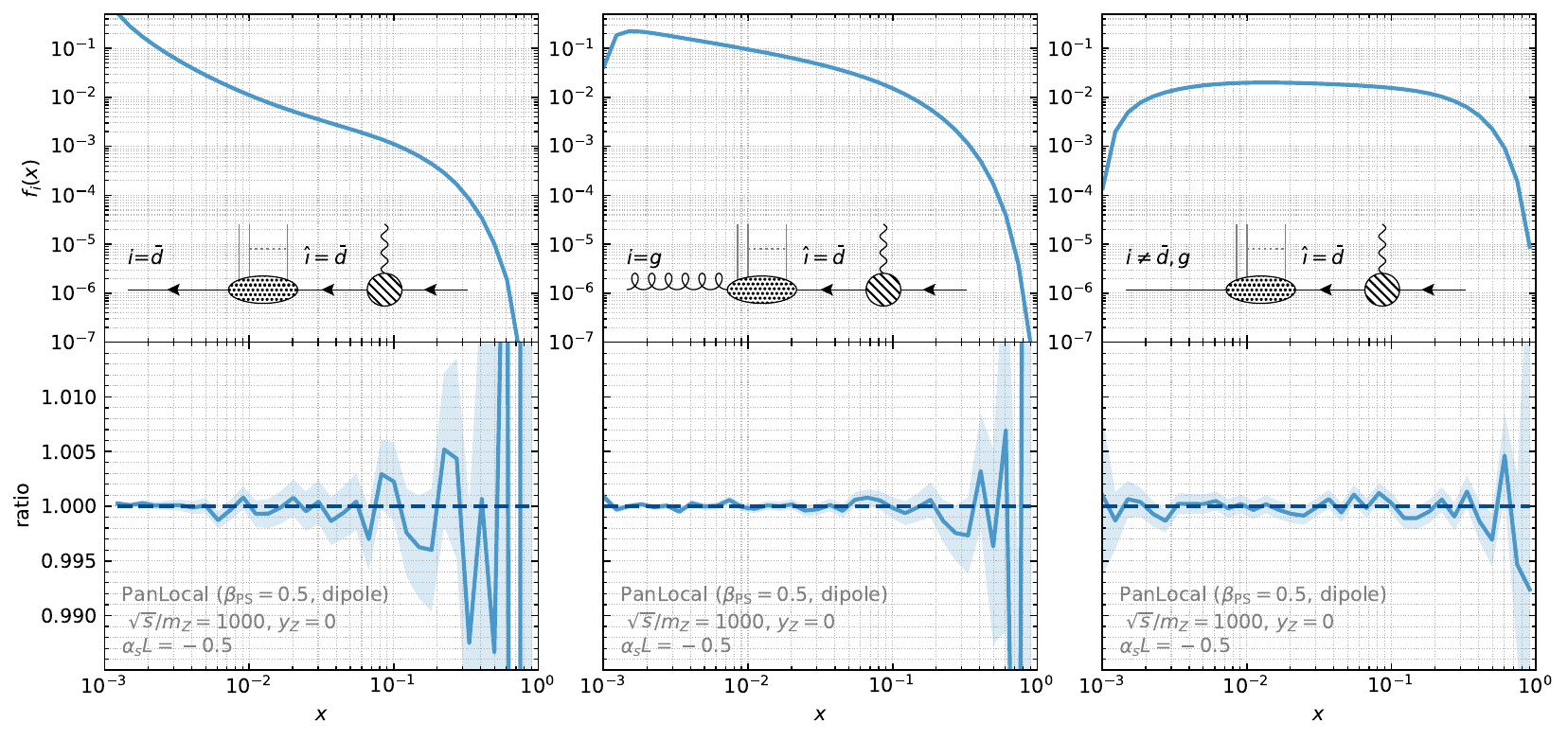}
	\end{subfigure}
	\caption{Ratio of the DGLAP evolution produced by the parton shower versus
		the DGLAP evolution as calculated with HOPPET.
		The results are
		shown as a function of the momentum fraction $x$
		carried by the parton extracted from the proton.
		The forward evolution in our HOPPET-based reference calculation is constrained to end with the
		$\ifinal=\bar{d}$ flavour, such that it reflects the starting
		point of the (backwards-evolving) shower,
		which we take to be $d\bar{d}$ in this case.
		We work with $y_Z=0$ and $\sqrt{s}/m_Z = 1000$, such that the maximal
		$x$ fraction a parton can have is $0.001$.
		We take $\lambda = \as L = -0.5$.  The three columns then show
		different extracted flavours $i$ that led to this 
		$\ifinal = \bar{d}$
		state, where we focus on the $i=\bar{d}$ (left), $i=g$ (middle)
		and $i\neq \{\bar d,g\}$ (right)
		cases.
		We show the PanGlobal shower with $\betaps = 0$ (upper panels), and
		the PanLocal dipole shower with $\betaps = 0.5$ (lower panels).  }
	\label{fig:pdf-origin-qqbar-1000GeV-beta0}
\end{figure}

For the purpose of comparing the shower to the DGLAP expectation we
are free to use either Eq.~(\ref{eq:x-origin-MC}) or the direct
shower-record information on the incoming parton $x$ after showering.
We choose the latter because it also gives easy access to the flavour
information.
We show results with a tiny value of $\as = 5 \times 10^{-6}$ and a
large value of $L = -10^5$, so as to render negligible any terms beyond
single-logarithmic accuracy.\footnote{In practice we use one-loop
  running of the coupling (rather than standard two-loop running), and
  do not include the $K_{\text{CMW}}$ two-loop cusp anomalous dimension.
  For purely single-logarithmic quantities, these choices have no
  impact on the results.
  Furthermore, in order to keep the event multiplicity under control,
  in the shower we discard radiation with a momentum fraction below
  some finite but small threshold $e^{-11}$, cf.\ Appendix~D of
  Ref.~\cite{Karlberg:2021kwr}. 
  In separate runs with a moderate value of $\as$, we have verified
  that such a cut does not impact the results.
  We use similar techniques also in
  Sections~\ref{sec:global-obs}--\ref{sec:rapidity-slice}, discarding
  radiation that will not affect the observable under study, again
  with a verification at finite $\alpha_s$ that this procedure does
  not affect the results.  }

We obtain the DGLAP prediction using the HOPPET evolution
code~\cite{Salam:2008qg}, which provides a straightforward way to
evaluate Eq.~(\ref{eq:DGLAP-origin}), as long as one ensures that
$\xorigin$ is not too close to $\xfinal$, to avoid systematic effects
associated with HOPPET's discretisation.
The HOPPET evolution is performed at single logarithmic accuracy,
i.e.\ leading-order (LO) evolution in the standard DGLAP nomenclature.
Since LO DGLAP evolution is purely single logarithmic, we are free to
use any finite $\as$ such that $\as L = -0.5$ as in the shower.
The treatment of PDFs, both in the shower and within HOPPET, is
further discussed in Appendix~\ref{sec:choice-pdfs}.
In particular, our approach for handling PDFs when working at very
small $\as$ values and large logarithms is discussed in
Appendix~\ref{sec:pdf-scale}, while the choice of PDFs at the
evolution starting scale is described in Appendix~\ref{app:pdfchoice}.

Results are shown in Fig.~\ref{fig:pdf-origin-qqbar-1000GeV-beta0}. 
We take $\sqrt{s}/m_Z = 1000$ and $y_Z = 0$ such that $\xfinal =
0.001$, and set $\ifinal = \bar{d}$.
We then consider three scenarios: the flavour of the incoming parton
remained the same ($\iorigin = \bar d$);
it became a gluon ($\iorigin=g$), which implies that at least one
flavour-changing splitting occurred;
or it became any other flavour ($i\neq \bar d,g$), which implies that
at least two flavour-changing splittings occurred.
The results are shown for the PanGlobal $\betaps = 0$ and the PanLocal
(dipole) $\betaps = 0.5$ showers (similar results are obtained for the
other showers, including both IF-recoil options of Dipole-$k_t$).
\logbook{}{see
  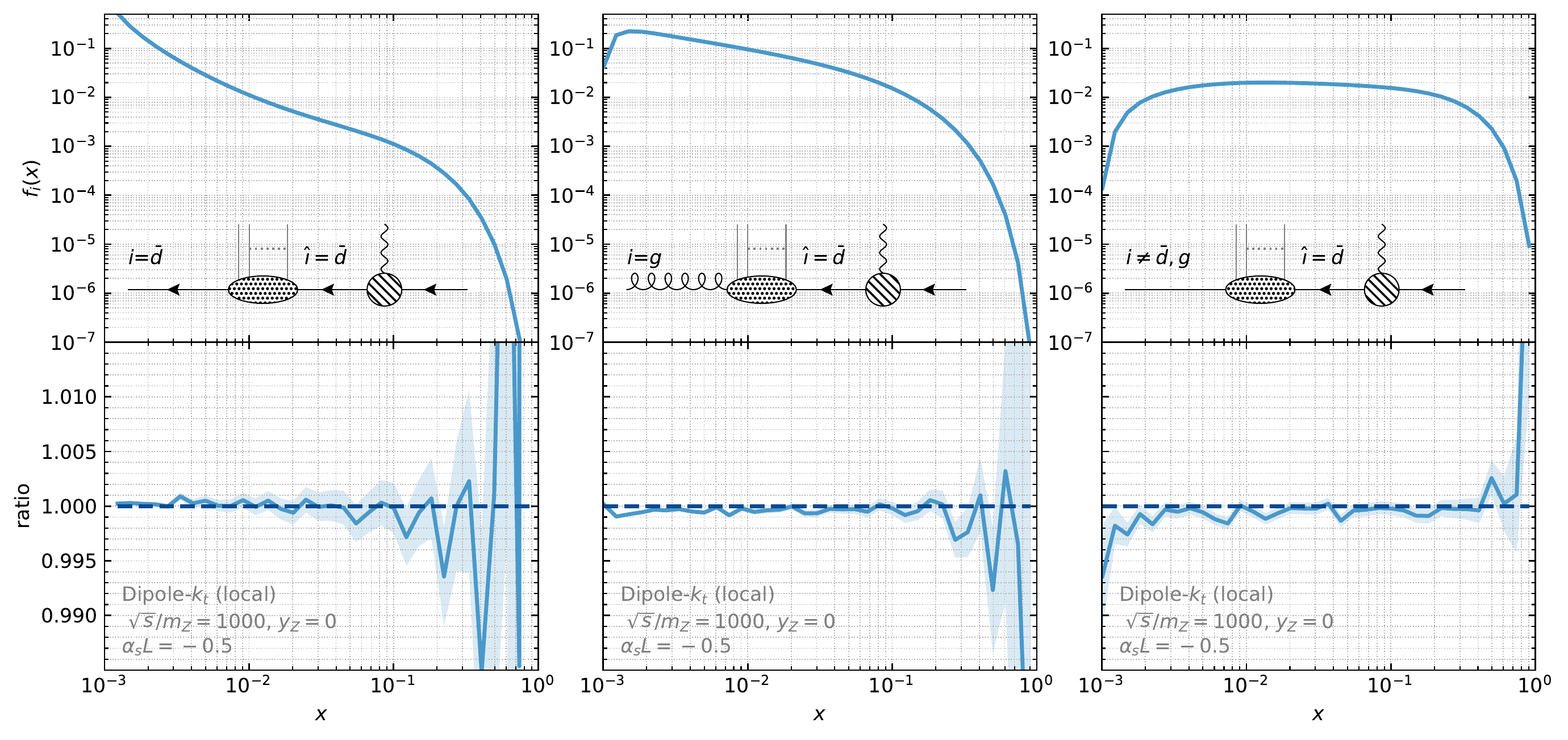
for the full set of showers tested}%
We obtain agreement with the predictions of standard DGLAP evolution
(with LO evolution, i.e.\ NLL accuracy in our context) to within
statistical accuracy.\footnote{This is less trivial than it sounds,
  both in terms of verifying the correctness of the implementation,
  and in terms of the interplay discussed in the past with the choice of ordering
  variable~\cite{Dokshitzer:2008ia,Skands:2009tb}.
}
The size of the statistical error depends on the value of the PDFs,
and is below $0.1\%$ in a substantial part of the $x$ range, but
increases in regions where the PDF is small or the flavour in question
is accessible only in rare events.

\begin{figure}[t]
    \centering
    \begin{subfigure}{\textwidth}
        \includegraphics[page=3, width=\textwidth]{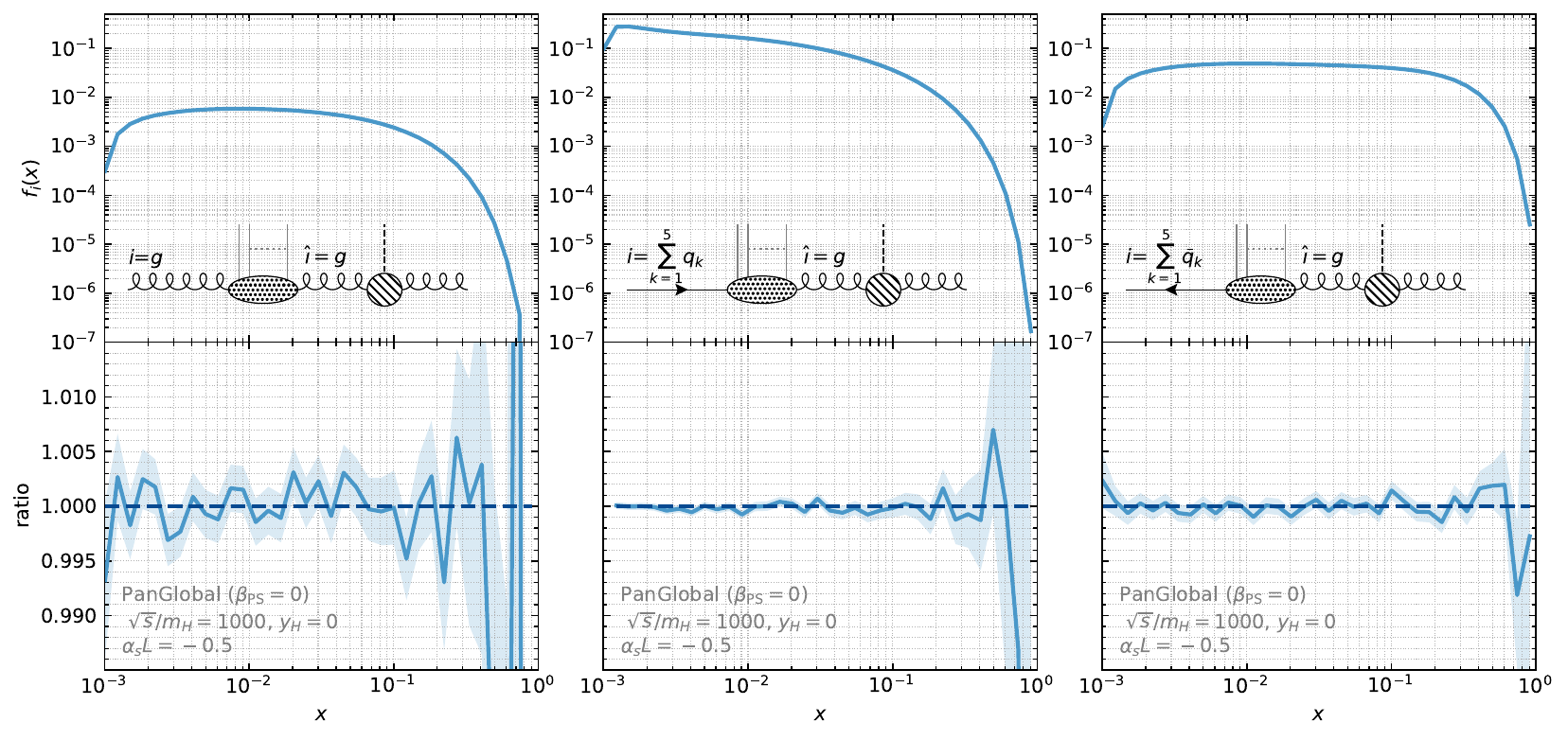}
    \end{subfigure}
    
    \begin{subfigure}{\textwidth}
        \includegraphics[page=5, width=\textwidth]{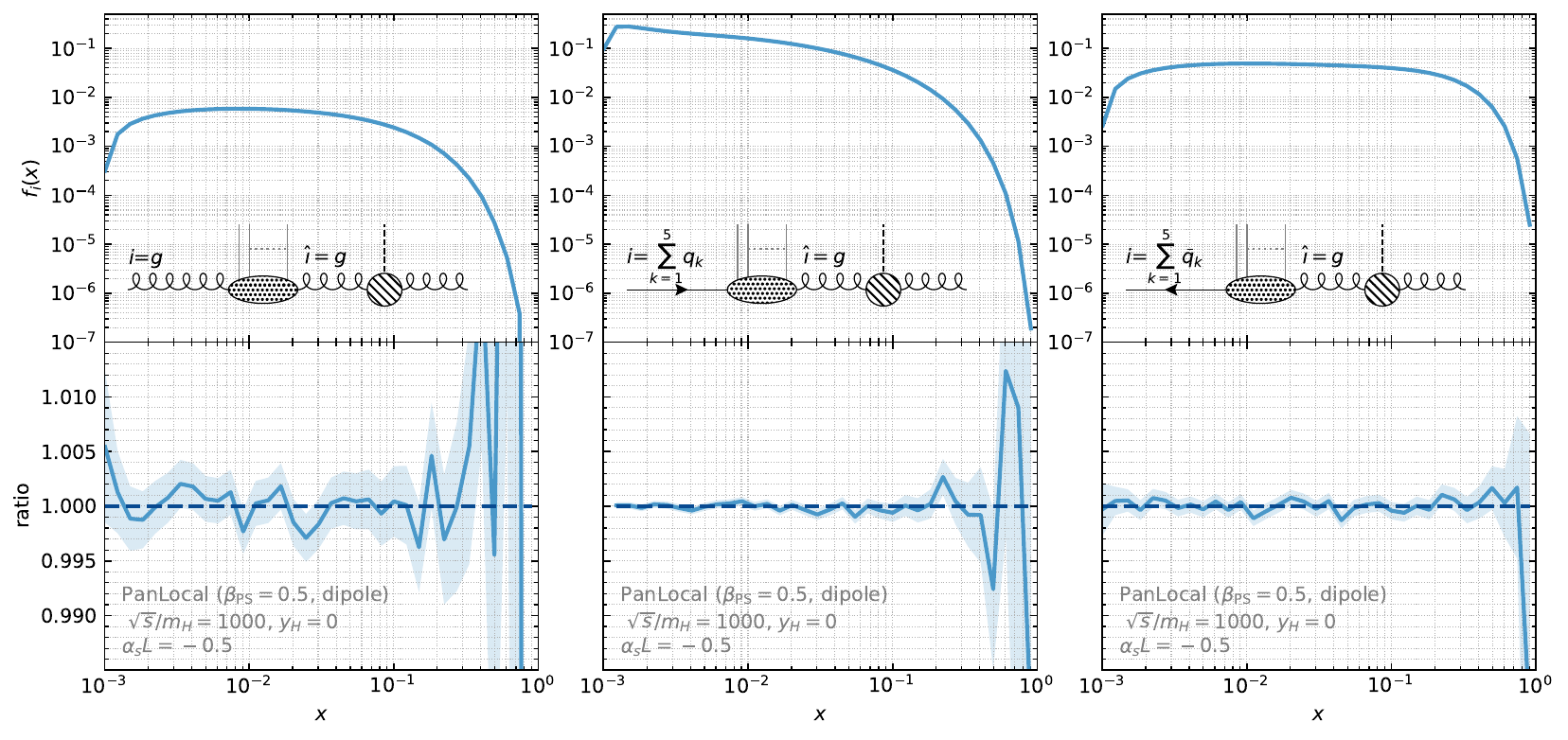}
    \end{subfigure}
    \caption{Same as Fig.~\ref{fig:pdf-origin-qqbar-1000GeV-beta0} but with the
    $gg\rightarrow H$ process. }
    \label{fig:pdf-origin-gg-1000GeV-beta0-lambda05}
\end{figure}

For completeness Fig.~\ref{fig:pdf-origin-gg-1000GeV-beta0-lambda05}
shows the distributions for $gg\rightarrow H$ events with
$\sqrt{s}/m_H = 1000$ and $y_H = 0$, where we examine $\iorigin = g$,
the sum over quarks $\iorigin = \sum_k q_k$, and the sum over
anti-quarks $\iorigin = \sum_k \bar{q}_k$.
Again, the agreement is good to within statistical errors, both for
the PanScales showers (shown) and the standard Dipole-$k_t$ showers
(not shown).
We have also tested different values of $\lambda$ and
$\sqrt{s}$. 
\logbook{}{see other PDF files in 2020-eeshower/analyses/pp-analyses/results/pdf-origin/}

\section{NLL tests for global observables}
\label{sec:global-obs}

A range of important collider observables belong to the class of
``global observables'', so-called because they are sensitive to
radiation in the whole of phase space.
These observables vanish in the absence of any radiation.
They include phenomenologically important quantities such as the
colour-singlet transverse momentum and the leading-jet transverse
momentum.
It is therefore of critical importance to understand the logarithmic
accuracy of showers for these observables.

Global observables share the feature that the probability (or
cumulative distribution) for a (dimensionless) observable $O$ to take
a value smaller than ${\rm e}^L$ can be written as~\cite{Catani:1992ua,Banfi:2004yd}
\begin{equation}
    \Sigma(O< e^L)\equiv \Sigma(\as,\as L) = H(\as)\exp\left[-L g_1(\as L) + g_2(\as L) 
    +\mathcal{O}(\as^n L^{n-1})\right] + \dots,
    \label{eq:NLLcumulative}
\end{equation}
where the ellipses denote corrections that are suppressed by powers of
$e^L$ (recall that $L$ is large and negative).
The function $H(\as)$ is the hard function multiplying the resummed series
and we shorten $\as(m_X^2)\equiv \as$. One may take $H(\as ) = 1$ at NLL accuracy.
In general, the N$^k$LL function $\as^{k-1}g_{k+1}(\as L)$ resums terms of
$\as^n L^{n-k+1}$.
In order to validate the NLL accuracy of the shower, we examine the
ratio of the parton shower evaluation of $\Sigma$ to the analytic NLL
evaluation, and check whether that ratio converges to 1 when one
extrapolates $\alpha_s\to 0$.

In this section we concentrate on observables measured on the hadronic
final state.
Given its particular phenomenological importance and subtle analytic
resummation properties, the discussion of the transverse momentum of
the $Z/H$ boson is deferred to Section~\ref{sec:transverse-mom}.
All of the tests here use $\sqrt{s}=5m_X$ and $y_X=0$.
We have also carried out a number of tests with $y_X=2$, which give
identical results, so we do not display them here.

\subsection{Leading jet transverse momentum and the azimuthal difference between
the two leading jets}
\label{sec:max-jet-pt}
We start by considering the transverse momentum, $p_{t1}$, of the hardest jet in
the colour-singlet production the process.
The quantity $\Sigma(p_{t1})$ corresponds to the efficiency of a jet
veto in colour-singlet production processes --- recall that jet vetoes
are widely used to reduce backgrounds to Higgs and other electroweak
production processes (e.g.\ backgrounds with leptons and missing
energy from top-quark production, which inevitably also involve jets).
Here we consider jets defined with the Cambridge/Aachen (C/A) algorithm with 
$R=1$~\cite{Dokshitzer:1997in,Wobisch:1998wt}, keeping in mind that the NLL
prediction is independent of the jet radius $R$ and is the
same~\cite{Banfi:2012yh} for all members of the generalised-$k_t$
family, including the anti-$k_t$ algorithm~\cite{Cacciari:2008gp}.
In Fig.~\ref{fig:maxpt-deltaphi-a} we show the
$\as \to 0$ extrapolation for the ratio of the shower cumulative
distribution to the NLL result, for the $pp\to Z$ process.
We see that the PanScales showers reproduce the analytic answer, i.e.\ 
$\lim_{\alpha_s\to 0}\Sigma_\text{PS}/\Sigma_\text{NLL}=1$.
That is not the case for Dipole-$k_t$ showers, with discrepancies of
up to $20\%$ for global IF recoil at extreme values of $\lambda$, and
$25\%$ with the local IF recoil Dipole-$k_t$ variant.

One comment is that these significant effects are in a region
corresponding to jet transverse momenta of the order of a few GeV, which is much
smaller than typical jet veto scales.\footnote{For example, in $H \to
  WW$ studies, it is common to use a $30\GeV$ jet veto, which
  translates to $\lambda \simeq -0.16$.
  Jet vetoes are also used in slepton (e.g.\ Ref.~\cite{CMS:2020bfa}) and
  electroweakino (e.g.\ Ref.~\cite{ATLAS:2021moa}) searches, with $\lambda$ values
  reaching of the order of $-0.3$.  \logbook{}{Looking at that ref,
    2106.01676, $SR^{WZ}-8$ has $MET>350\GeV$, which implies
    $\sqrt{\hat s} > 700\GeV$, and it involves a jet veto with a
    $20\GeV$ scale, so $\lambda = -0.326$.  } }
However, jet activity at such low momenta is relevant also in studies
of multiple interactions and the underlying
event~\cite{CMS:2017ngy,ATLAS:2019ocl}. 
Specifically, underlying event studies often examine energy and
charged-particle flow in different azimuthal regions of the event,
defined with respect to the $Z$ transverse direction.
Another context in which azimuthal correlations are important is in
the identification of ridge-like structures in high-multiplicity $pp$
collisions~\cite{CMS:2010ifv,ATLAS:2015hzw}.

\begin{figure}[tb]
	\centering
	\begin{subfigure}{0.48\textwidth}
		\includegraphics[page=1,height=0.3\textheight]{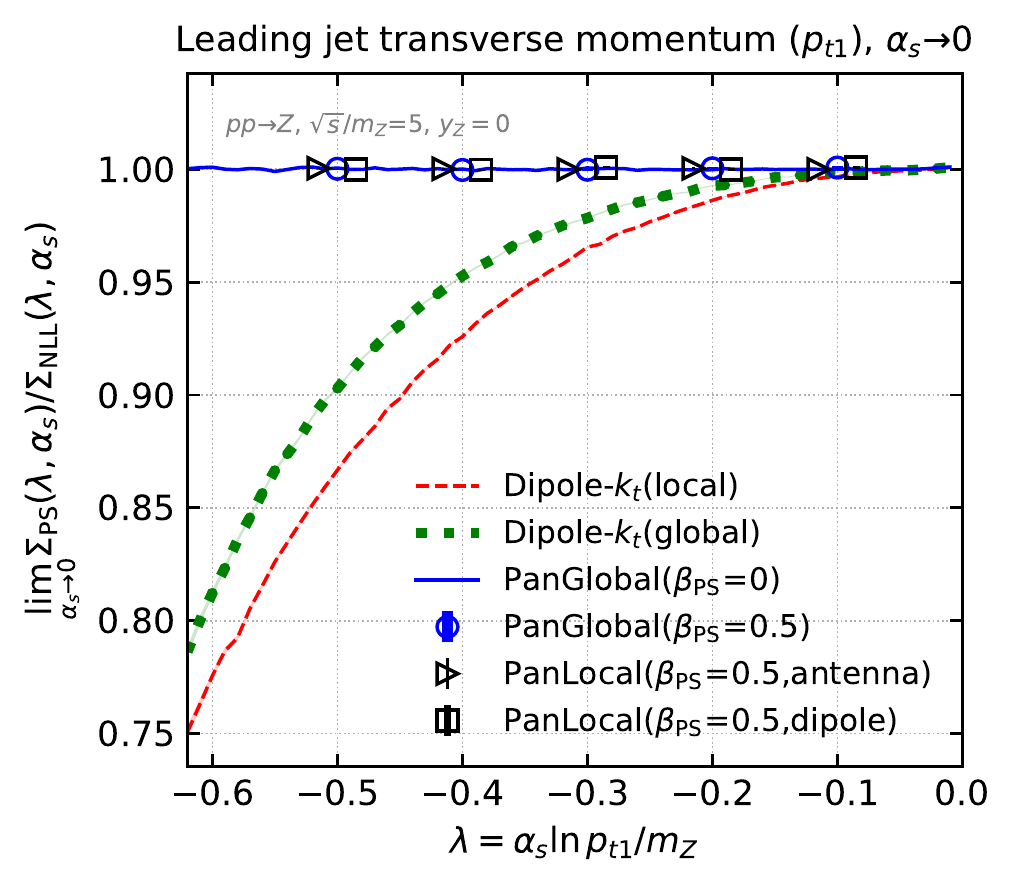}
		\caption{}
		\label{fig:maxpt-deltaphi-a}
	\end{subfigure}
	\hfill
	\begin{subfigure}{0.48\textwidth}
		\includegraphics[page=1,height=0.3\textheight]{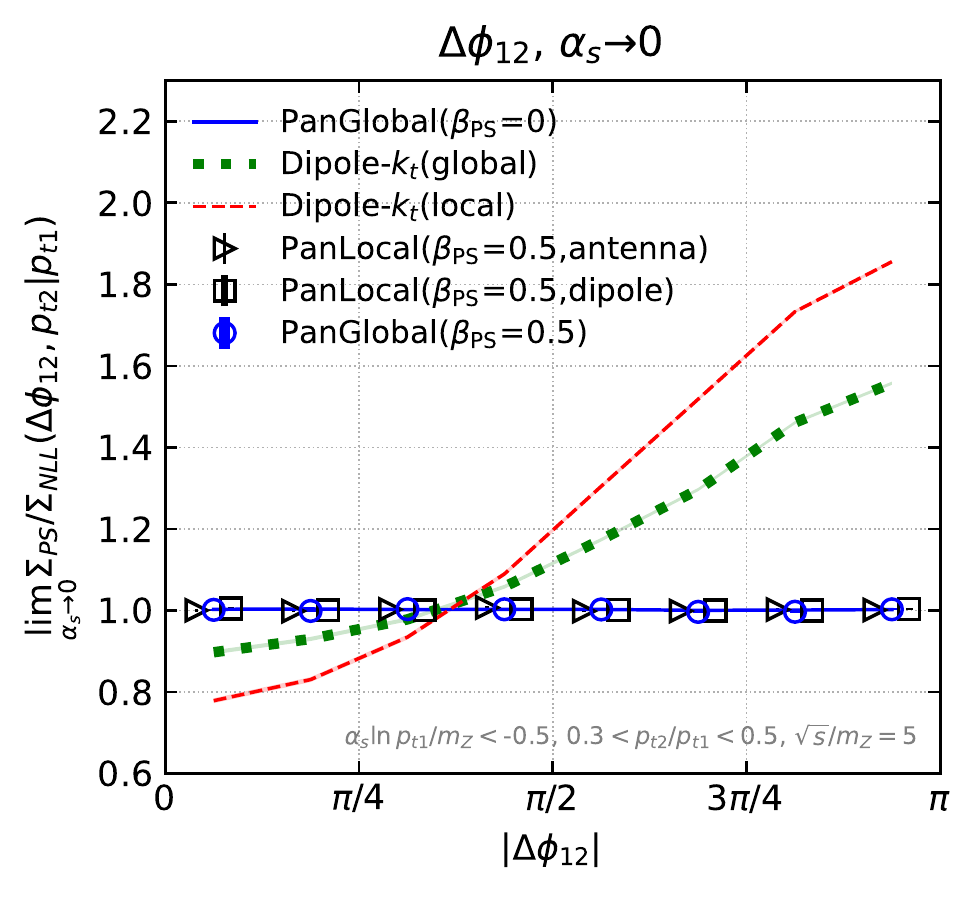}
		\caption{}
		\label{fig:maxpt-deltaphi-b}
	\end{subfigure}
	\caption{ (a) Ratio of the shower and NLL results for the cumulative
		distribution for the hardest jet transverse momentum, $p_{t1}$, with $-0.62 < \lambda <
		0$ and $\as \to 0$ for $pp\to Z$ events.
		(b) Difference in azimuthal angles
		between the two leading jets in a $pp\to Z$ event, where the first jet has
		$\lambda = \as \ln p_{t1}/m_Z \leq -0.5$, while the second one has $0.3 \leq
		\frac{p_{t2}}{p_{t1}} \leq 0.5$.
		In the $\as \to 0$ limit, the $\lambda<-0.5$ condition effectively
		fixes $\lambda=-0.5$.
		In both plots, we have shifted the horizontal locations of the
		markers for the two PanLocal showers, so as to avoid having
		all the symbols overlap. 
	}
	\label{fig:maxpt-deltaphi}
\end{figure}

To give some insight into possible azimuthal structures induced by
parton showers, we study a specific observable, namely the
distribution of the difference in azimuthal angles between the two
highest-$p_t$ jets, $\Delta \phi_{12}$.
At NLL, this distribution is flat in $|\Delta \phi_{12}|$ and reads
\begin{equation}
    \Sigma(\Delta \phi_{12},p_{t2}|p_{t1}) \equiv \frac{\Sigma(\Delta 
      \phi_{12},p_{t2},p_{t1})}{\Sigma(p_{t1})}
    =
    \frac{1}{\pi}\left(e^{2 C_i 
    R_0^\prime(-b_0\lambda)\ln p_{t2}^{\max}/p_{t1} } - e^{2 C_i R_0^\prime(-b_0\lambda)\ln 
    p_{t2}^{\min}/p_{t1} }\right),
    \label{eq:deltaphinll}
\end{equation}
with the $R_0'$ function as given in Eq.~(\ref{eq:Rp-beta0}). 
Fig.~\ref{fig:maxpt-deltaphi-b} shows the $\as \to 0$ limit of that
distribution, normalised to the NLL result, for events where
$\as \ln p_{t1}/m_Z < -0.5$ and $0.3 < p_{t2}/p_{t1} < 0.5$.
Again we see that the PanScales showers reproduce the NLL expectation.
The Dipole-$k_t$ showers do not, with up to 85\% (55\%) discrepancies
when a local (global) IF recoil is employed,
a consequence of the way in which they perform the transverse momentum
recoil.
Note that in the $\as \to 0$ limit, the two jets that are relevant for
Fig.~\ref{fig:maxpt-deltaphi-b} are nearly always both soft and
well separated in rapidity.
Consequently, at NLL accuracy, the observable is not affected by spin
correlations. 

A final comment in this section is that the NLL discrepancies that we
observe for the IF-global Dipole-$k_t$ variant are expected (and
observed) to be the same as those for related observables in $e^+e^-$
collisions~\cite{Dasgupta:2020fwr} (modulo the fact that the latter's
results were at leading colour, while here we use the NODS colour
scheme~\cite{Hamilton:2020rcu}).\footnote{For the leading jet $p_t$ in
  Fig.~\ref{fig:maxpt-deltaphi-a}, the discrepancy at $\lambda=-0.5$
  agrees with what was found for $M_{\beta=0}$ in Fig.~11 of
  Ref~\cite{Hamilton:2020rcu}.}
Indeed, the choice of the evolution variable, as well as the dipole
being partitioned in its rest frame, is common to both initial- and
final-state formulations, at least in the soft-and-collinear limit
relevant for these NLL discrepancies.

\subsection{Generic global event shapes}

\label{sec:generic-global}

Next, we discuss a wider range of global event shape observables.
For this purpose, it is useful to introduce three families of observables:
\begin{subequations}
    \label{eq:global-obs}
  \begin{align}
    S_{p, \beta_\text{obs}} &= \!\!\!\!
                              \sum_{i \in \text{partons}} \frac{p_{t i}}{Q} {\rm e}^{-\beta_{\rm
                             obs} |y_i - y_X|}\,,\\
    S_{j, \beta_\text{obs}} &= \sum_{i \in \text{jets}} \frac{p_{t i}}{Q} {\rm e}^{-\beta_{\rm
                             obs} |y_i - y_X|}\,,\\
    \label{eq:global-obs-Mj}
    M_{j, \beta_\text{obs}} &= \max_{i \in \text{jets}} \frac{p_{t i}}{Q} {\rm e}^{-\beta_{\rm
                             obs} |y_i - y_X|}\,,
  \end{align}
\end{subequations}
where $p_{ti}$ and $y_i$ are respectively the transverse momentum and
rapidity of parton or jet $i$, $y_X$ is the rapidity of the
colour-singlet system, and jets are again defined with the C/A
algorithm with $R=1$.
The ``$S$'' observables involve a sum over either particles or jets,
while the ``$M$'' observables examine a maximum across jets.
Each family is parametrised by a variable $\beta_\text{obs}$, which
determines the relative weighting of central versus forward
particles/jets.
Note that $M_{j,0}$ coincides with the transverse momentum of the
hardest jet shown in Fig.~\ref{fig:maxpt-deltaphi-a},
while $S_{p,1}$ coincides with the widely studied $0$-jettiness
($\tau_0$) of Ref.~\cite{Stewart:2010tn}, which is also used in the
Geneva~\cite{Alioli:2012fc,Alioli:2013hqa} matching procedure.
For all of the observables, the LL resummation structure depends on
$\beta_\text{obs}$.
For a given value of $\beta_\text{obs}$, the $M_{j, \beta_\text{obs}}$
observables differ from the $S_{p, \beta_\text{obs}}$ and
$S_{j, \beta_\text{obs}}$ at NLL, while the $S_{p, \beta_\text{obs}}$
and $S_{j, \beta_\text{obs}}$ observables differ from NNLL onwards.
The resummation formulas up to NLL are summarised in
Appendix~\ref{app:resummation}.

In our numerical tests, we take $\beta_\text{obs} = 0,0.5, 1$.
In Fig.~\ref{fig:global-summary} we show the ratio of the shower to
the NLL result for the cumulative distribution $\Sigma(O<e^L)$, as
calculated in the limit $\as \to 0$ for $\lambda=-0.5$.
As in the final-state case~\cite{Dasgupta:2020fwr}, we find that
standard dipole showers fail to reproduce the all-order NLL results for
$\beta_{\text{obs}}=0$ observables, as represented by the red squares.
This failure is a consequence of incorrect assignment of transverse
recoil to earlier emissions~\cite{vanBeekveld:2022zhl}.
Its impact on logarithmic terms can be examined analytically with a
fixed-order study analogous to that in the final-state
case~\cite{Dasgupta:2018nvj}.
Concerning the $\beta_{\text{obs}}=0.5,1$ cases, the $\as \to 0$
dipole-shower results appear to agree with the NLL predictions.
However the studies of similar observables in the final-state case
showed that dipole-type showers induce spurious all-order
leading-colour super-leading logarithms, $(\as L)^n (\as L^2)^p$
(Section 2-d of the supplementary material of
Ref.~\cite{Dasgupta:2020fwr}).
Because these issues arise from the soft-collinear region, which is
effectively treated identically in the final-state and (global-IF)
initial-state cases, they will inevitably arise also in the
initial-state case (for local-IF recoil, we expect similar problems).
Accordingly we colour these dipole-shower points in amber.
The green circles for the four PanScales showers in
Fig.~\ref{fig:global-summary} indicate that their predictions are in
agreement with the NLL results, and the analysis of recoil in
Ref.~\cite{vanBeekveld:2022zhl} ensures the absence of the fixed-order
issues that cause us to colour the dipole showers in amber.

As a final remark, we remind the reader that in these studies,
subleading $N_c$ corrections have been included according to the NODS
method~\cite{Hamilton:2020rcu} for both the dipole-type showers and
the PanScales showers, so as to concentrate on the impact of
recoil.
In contrast, standard dipole showers choose the colour factor
according to whether the emitting dipole end that is closer (in the
dipole centre-of-mass frame) is a gluon ($C_A/2$) or a quark ($C_F$).
This results in incorrect terms already at LL, in analogy with the
final-state discussion in Ref.~\cite{Dasgupta:2018nvj}.
The numerical impact will be the same as in the all-order final-state
study~\cite{Hamilton:2020rcu}.

\begin{figure}[tb]
	\centering
	\includegraphics[page=2,width=0.9\textwidth]{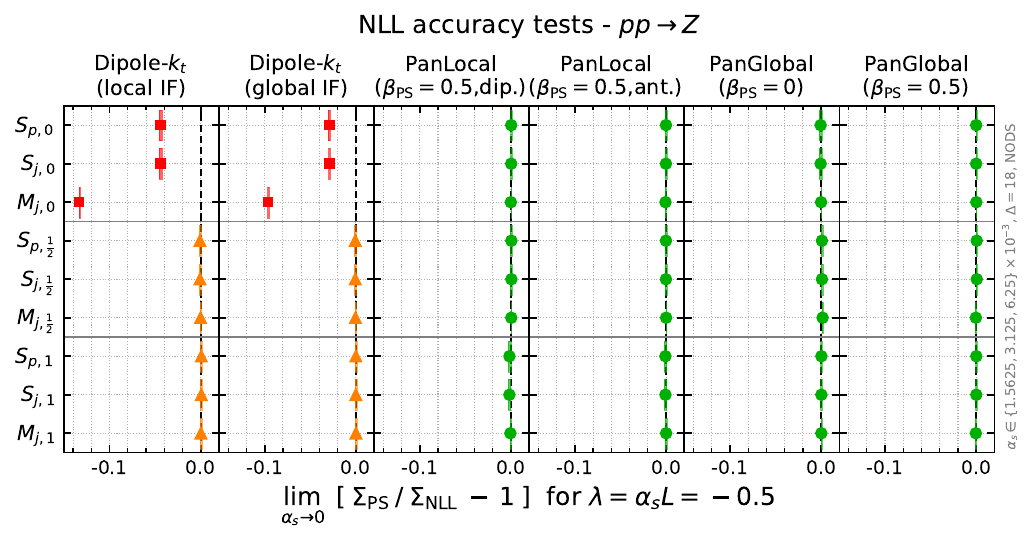}
	\caption{ Summary of deviations from NLL for several global observables for
		the process $q\bar{q}\to Z$ and $\lambda=-0.5$. Red squares denote a clear
		NLL failure; amber triangles indicate a NLL fixed-order failure that is
		masked at all orders; green circles are used when the shower
		passed both the numerical NLL tests and the fixed-order recoil tests.
		The $\as\to0$ result is obtained by quadratically extrapolating the shower
		results at $\as =0.00625, 0.003125$ and $0.0015625$, and includes
		a systematic error that is evaluated as the change in the $\as \to
		0$ extrapolation when one
		uses $\as=0.0125$ instead of $\as=0.003125$.
		The showers include a dynamic cutoff $\Delta=18$, which functions
		as discussed in our earlier $e^+e^-$ tests~\cite{Dasgupta:2020fwr,Hamilton:2020rcu}.
	}
	\label{fig:global-summary}
\end{figure}
\section{The transverse momentum of the colour-singlet system}
\label{sec:transverse-mom}

The next observable that we discuss is the cumulative distribution for
the transverse momentum of a massive colour singlet (here, $Z$ or $H$
boson) produced in proton collisions.
It has wide relevance for LHC phenomenology, and for example its
understanding is critical for $W$ mass
extractions~\cite{ATLAS:2017rzl,LHCb:2021bjt,CDF:2022hxs}.\footnote{One
  should keep in mind, that in many applications parton showers are
  reweighted so that the colour-singlet transverse momentum
  distribution agrees with high-order matched resummed and fixed order
  predictions, such as~\cite{Bizon:2019zgf,Alioli:2021qbf,Re:2021con,
  Becher:2020ugp,Camarda:2021ict,Billis:2021ecs,Ebert:2020dfc,
  Chen:2018pzu,Chen:2022cgv,Ju:2021lah,Neumann:2022lft}. 
  Still, even if such a procedure results in a correct colour-singlet
  transverse momentum distribution for the reweighted shower, it will
  not in general correctly account for correlations between the colour
  singlet and the full pattern of hadronic energy deposition.
  We leave the detailed study of such questions to future, more
  phenomenological work.  }
It is also widely used in matching showers and fixed-order
calculations~\cite{Hamilton:2012rf,Monni:2019whf,Buonocore:2022mle,Alioli:2021qbf}.

The colour singlet $p_{tX}$ distribution is a more subtle observable
than those studied in the previous subsections, essentially because it
has two resummation regimes.
In one of the regimes, that with moderately small $p_{tX}$, the
suppression of the cross section is driven dominantly by the Sudakov
suppression of emissions and the NLL prediction can be written in
terms of our standard resummation formula,
Eq.~(\ref{eq:NLLcumulative}), where $L = \ln p_{tX}/m_X$.
The other regime concerns asymptotically small values of $p_{tX}$,
which are typically obtained by a vector cancellation between the
recoils from two or more gluons emitted with $p_t$'s substantially
larger than $p_{tX}$.
In this regime, Eq.~(\ref{eq:NLLcumulative}) breaks
down~\cite{Frixione:1998dw}, and the NLL resummation instead generally
requires $b$-space resummation~\cite{Parisi:133268},\footnote{A direct
  $p_t$-space solution to this issue is presented in
  Ref.~\cite{Monni:2016ktx}.} giving a result for $\Sigma$ that scales
as $p_{tX}^2$.
The transition between the two regimes occurs where
$R' = |\partial_L (L g_1(\as L))| = 2$ and it is reflected in a
$1/(R'-2)$ divergence in the $g_2(\as L)$ function of
Eq.~(\ref{eq:NLLcumulative}), cf.\ Eqs.~(\ref{eq:g2-nll}) and
(\ref{eq:curlyF_Vpt}) of Appendix~\ref{app:resummation}.
The location of the transition corresponds to $\lambda \simeq -0.48$
for $Z$ production and $\lambda \simeq -0.32$ for Higgs production (both for $n_f =
5$).
In the region of moderately small $p_{tX}$ (``Sudakov region'') we
will carry out our tests in the same way as earlier, while for
asymptotically small $p_{tX}$ (``power-scaling region'') we will
adopt a somewhat different procedure.
We start with the former.

\subsection{Sudakov region}
\begin{figure}[t]
  \centering
  \begin{subfigure}{0.48\textwidth}
    \includegraphics[page=1,width=\textwidth]{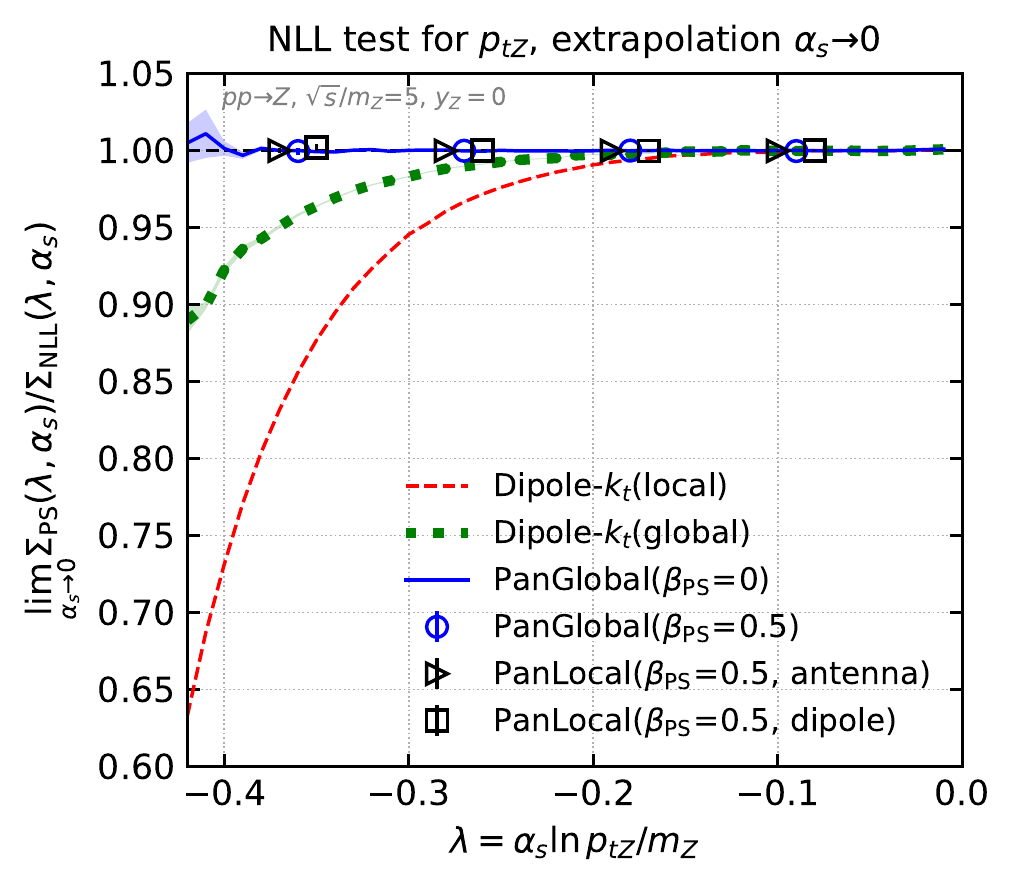}
    \caption{}
    \label{fig:ZHpt-distribution-a}
  \end{subfigure}
  \hfill
  \begin{subfigure}{0.48\textwidth}
    \includegraphics[page=1,width=\textwidth]{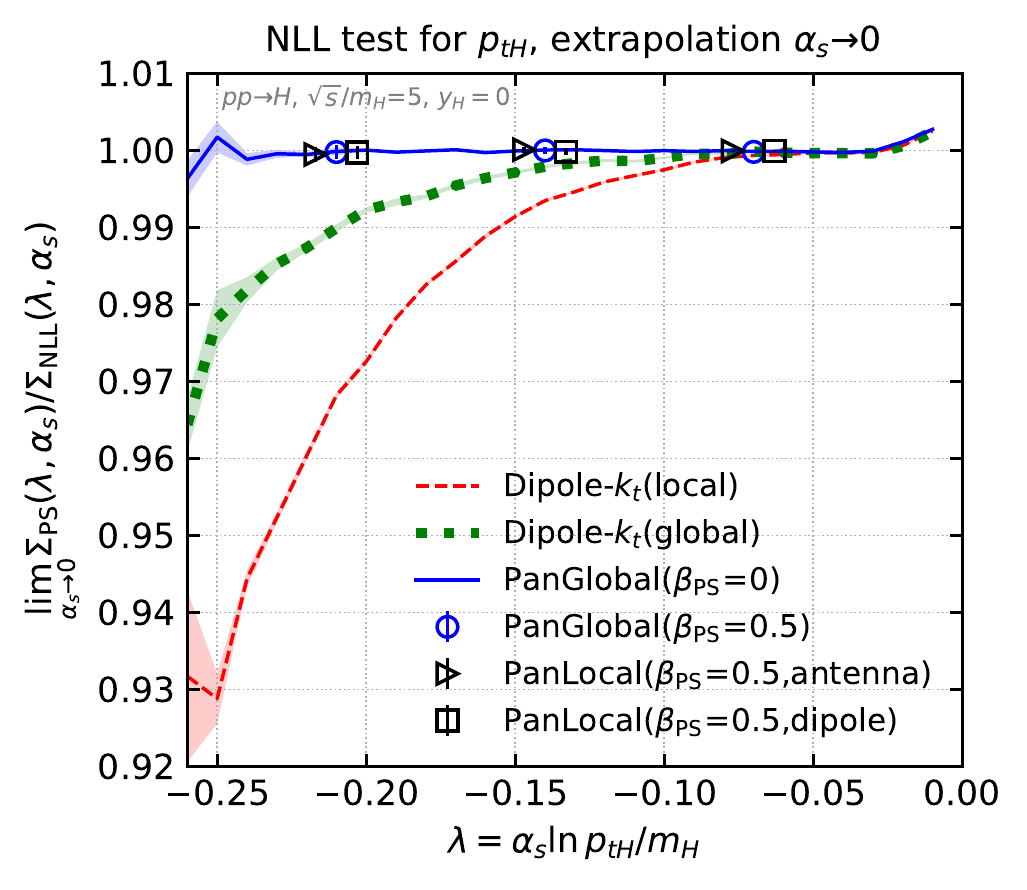}
    \caption{}
    \label{fig:ZHpt-distribution-b}
  \end{subfigure}
  \caption{Ratio of the cumulative distribution for the colour-singlet
    transverse momentum to the NLL analytic result, in
    the $\as\to 0$ limit, 
    for (a) $q\bar q\to Z$ and  (b) $gg\to H$  events.
    The results are shown for
    Dipole-$k_t$ with local (red dashed line) and global recoil (green dotted
    line), PanGlobal with $\beta_{\text{PS}}=0$ (blue solid line) and
    $\beta_{\text{PS}}=0.5$ (blue circles), and PanLocal with
    $\beta_{\text{PS}}=0.5$, both for the antenna (black triangles) and dipole
    (black squares) variants. For clarity, the PanLocal antenna
    (dipole) points have been slightly shifted towards the left (right), with
    respect to the values actually used, which coincide with the PanGlobal
    $\betaps=0.5$ ones.
    }
  \label{fig:ZHpt-distribution}
\end{figure}

In Fig.~\ref{fig:ZHpt-distribution} we show the $\as \to 0$
extrapolation of the ratio of the shower to the NLL prediction for the
colour singlet transverse momentum, with a range of showers.
The results are shown for $q\bar{q}\to Z$
(Fig.~\ref{fig:ZHpt-distribution-a}) and $gg\to H$
(Fig.~\ref{fig:ZHpt-distribution-b}). 
We use Eq.~(\ref{eq:NLLcumulative}) as our NLL
reference together with ingredients from Eqs.~\eqref{eq:g1-b0},
\eqref{eq:g2-nll} (using $\beta_{\text{obs}}=0$), and
Eq.~\eqref{eq:curlyF_Vpt}.
We consider only $\lambda \ge -0.42$ ($Z$) and $\lambda \ge -0.26$ ($H$), to
stay well away from the breakdown of the $p_{tX}$-space NLL
resummation.
The PanScales showers that are shown all agree with the NLL
prediction.
Conversely, the Dipole-$k_t$ showers fail to reproduce the correct NLL
result.
For the $q\bar{q}\to Z$ process, we see a 35\% (10\%) discrepancy of
the NLL terms at $\lambda=-0.42$ using the Dipole-$k_t$ shower with a
local (global) recoil.  For the $gg\to H$ process we find a 7\% (3\%)
difference at $\lambda = -0.26$.
Performing a comparison at the same value of $\lambda = -0.25$ for both
processes, we find a $3\%$ ($1\%$) discrepancy for $q\bar{q}\to Z$
versus $7\%$ ($2\%$) for $gg\to H$, with local (global) IF recoil in
the Dipole-$k_t$ shower.\footnote{%
  One can develop an intuition for the sizes of the effects across
  different observables and different processes with the help of a
  fixed-order analysis of the kind carried out in
  Ref.~\cite{Dasgupta:2018nvj}.
  Many of the results from that article carry over to the
  initial-state case, however we leave a detailed analysis to the
  interested reader.  }

\subsection{Power-scaling region}

Now let us turn to the second resummation regime, namely that where $R' > 2$
and the dominant mechanism to produce a small $p_{tX}$ is a vector
cancellation between the transverse recoils of different emissions.
A first remark is that the tests shown in
Fig.~\ref{fig:ZHpt-distribution} already probe this mechanism, because
the NLL result is sensitive to it even in the regime of $R' < 2$,
through the $g_2$ function in Eq.~(\ref{eq:NLLcumulative}),
specifically the part in Eq.~(\ref{eq:curlyF_Vpt}).
Still, the $R'>2$ regime is conceptually important and it is therefore
of interest to explicitly examine the behaviours of different
showers. 

It is useful to recall the structure of the standard $b$-space result
for the resummation of the transverse-momentum
distribution~\cite{Parisi:133268,Collins:1984kg,Bozzi:2005wk},
\begin{eqnarray}
    \label{eq:pt-bessel-resum}
  \frac{{ \rm d} \Sigma}{{\rm d}p_{tX}^2} = \int_0^{\infty} \frac{{\rm d}b}{2}b
  J_0(b p_{tX})\Sigma_V(b_0/b)\,,
\end{eqnarray}
with $b_0 = 2{\rm e}^{-\gamma_E}$, $\Sigma_V$ the $b$-space resummed
distribution, and $J_0$ the Bessel function of the first kind and
order $0$.
Observe that for $p_{tX} \to 0$ the result tends to a non-zero
constant, whose value can be straightforwardly obtained by replacing
$J_0(b p_{tX}) \to 1$ in Eq.~(\ref{eq:pt-bessel-resum}).
Fig.~\ref{fig:pt-asymptotic} shows the small-$p_{tX}$ behaviour of the
distribution for $Z$ production, in four showers.
Three of them, PanGlobal, PanLocal and Dipole-$k_t$(global), indeed
tend to a non-zero constant.
In contrast the variant of Dipole-$k_t$ with local recoil for IF
dipoles tends to zero in this limit, i.e.\ it has the wrong scaling
behaviour.
This is because, after the first emission, the event consists of two
IF dipoles, and from that point onwards, no further transverse recoil
is taken by the $Z$ boson.
Therefore the only mechanism for $p_{tZ}$ to be small is Sudakov
suppression of the first emission, which is a much stronger
suppression than the vector cancellation.\footnote{ For processes such
  as $gg \to H$ with two II dipoles, one does recover the correct
  power-dependence of the scaling (i.e.\ the plateau), because the Higgs recoil induced by
  an emission off one II dipole can have a vector cancellation with
  recoil induced by an emission off the other II dipole.
  However the normalisation of the plateau is still expected to
  be wrong, as is the whole shape of the distribution for $\as L\sim 1$.
  \logbook{}{See \ttt{2020-eeshower/analyses/pp-analyses/boson-pt-results/boson-pt-v-norm-H.pdf} and \ttt{2020-eeshower/analyses/pp-analyses/boson-pt-results/plot-summary-boson-pt-asym-H.pdf} which illustrate $gg\to H$ for $\alpha_s=0.3,0.4$}
}

\begin{figure}[t]
    \centering
    \begin{subfigure}{0.50\textwidth}
      \includegraphics[page=1,width=\textwidth]{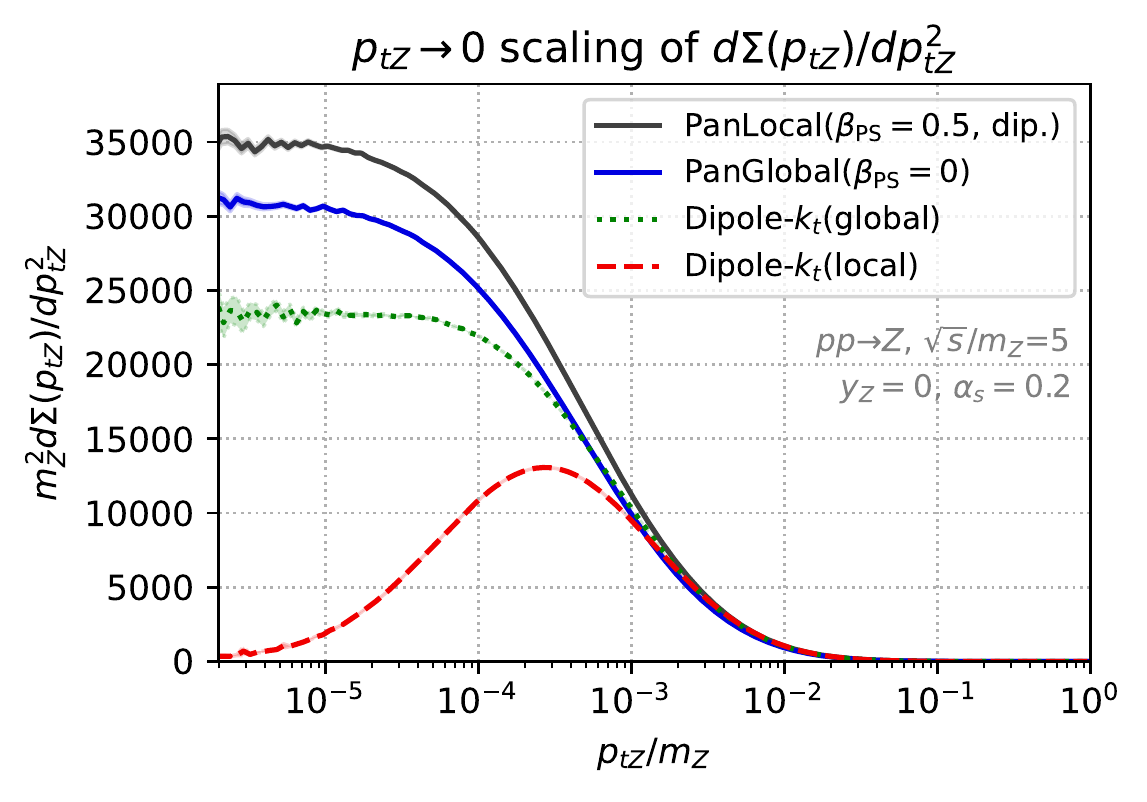}
      \caption{}
      \label{fig:pt-asymptotic}
    \end{subfigure}\hfill
    \begin{subfigure}{0.45\textwidth}
      \includegraphics[page=1,width=\textwidth]{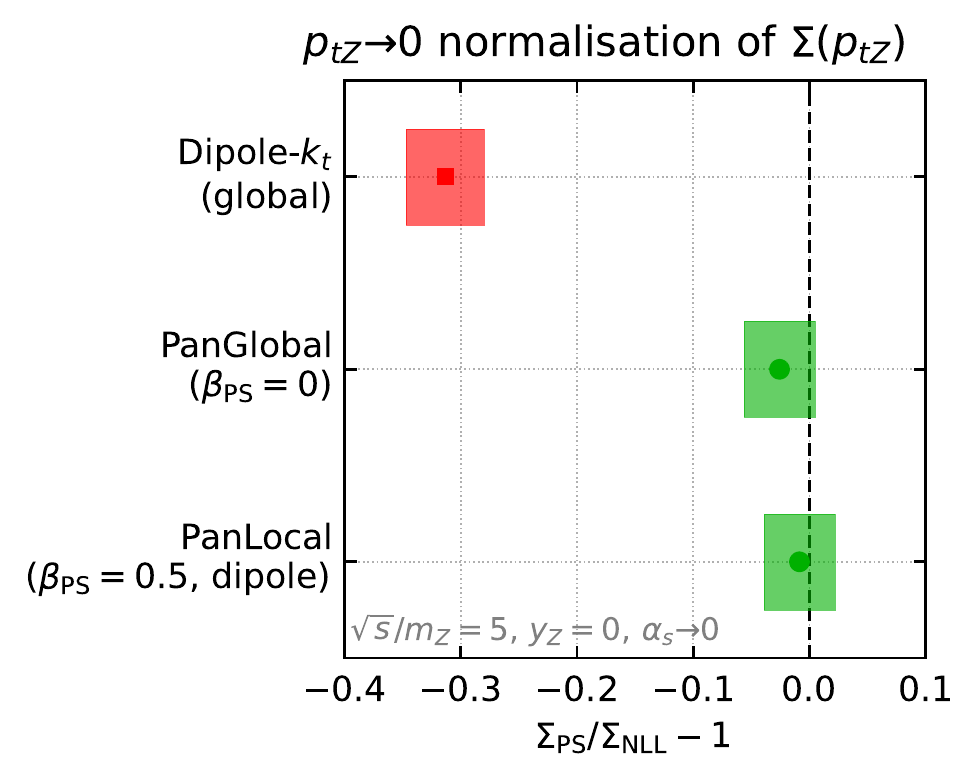}
      \caption{}
      \label{fig:as-to-0-pt-asym}
    \end{subfigure}
    \caption{(a) $m_Z^2  d\Sigma(p_{tZ})/d{p_{tZ}^2}$, as determined with four
  showers.
  In QCD this quantity tends to a
  calculable (non-zero) constant for $p_{tZ} \to 0$~\cite{Parisi:133268}.
  (b) For the three showers that tend to a non-zero
  constant, the plot shows the $\as \to 0$ limit of the deviation of
  that constant relative to the NLL expectation, with the usual 
  (red) green colour coding for (dis)agreement with NLL.
}
        \label{fig:ptZ-scaling}
\end{figure}

For those showers that do tend to a non-zero constant, it is worth
checking the value of that constant, which is a prediction of the NLL
resummation.
The expected value can be deduced from Eq.~(\ref{eq:pt-bessel-resum}),
simply setting $p_{tX}=0$ on the right-hand side.
Note that at our NLL accuracy, $\Sigma_V$ coincides with the
cumulative distribution of the leading jet $p_t$, or equivalently
(still at NLL), in a $p_t$-ordered shower, the shower ordering
variable.
We use the distribution of the latter (or a transverse-momentum like
analogue in $\betaps=0.5$ showers) to evaluate
Eq.~(\ref{eq:pt-bessel-resum}), because it facilitates the $\as \to 0$
extrapolation.

To determine the asymptotic normalisation of the shower, one needs to
evaluate the height of the plateau in Fig.~\ref{fig:pt-asymptotic}.
As can be seen in the plot, this is somewhat delicate because on one
hand the approach to the asymptotic value is fairly slow,\footnote{For
  example, with the setup of Fig.~\ref{fig:pt-asymptotic},
  $\alpha_s=0.2$, one would reach the transition point where $R'=2$ at
  $p_{tZ}/m_Z=\exp(-\pi/(2C_F\alpha_s))\approx 2.8\,10^{-3}$, which is
  almost two orders of magnitude larger than the observed plateau.
  With a running coupling we expect the transition between the
  two regimes to be more rapid.
}
and on the other hand the statistical errors grow rapidly at small
$p_{tZ}$.
For each value of $\as$ that we study, we estimate the ratio of the
shower plateau height to the NLL expectation in a $p_{tZ}$ region
where Eq.~(\ref{eq:pt-bessel-resum}) is within $3\%$ of its asymptotic
value, assigning as a systematic error the change induced when
increasing $\ln p_{tZ}$ by one.
\logbook{}{Watch out: we take the ratio to the asymptotic value for
  dipole-$k_t$ and to the Bessel transform for the PanScales showers.
  The different choices reflect the option that give the best $p_{tZ}
    \to 0$ convergence in each case
}%
We then perform a linear extrapolation of the $\as=0.2$ and $0.3$
ratios to obtain the ratio at $\as = 0$, with a further systematic
obtained from the change in the result when instead using $\as=0.2$
and $0.4$.
Finally, we account for the fact that the plateau is determined in a
region that is $3\%$ away from the asymptotic region with a further
overall $3\%$ systematic error (which ultimately dominates the total
error).
The final ratios, with total statistical and systematic errors are
shown in Fig.~\ref{fig:as-to-0-pt-asym}.
The PanGlobal ($\betaps=0$) and PanLocal ($\betaps=0.5$) showers are
consistent with the NLL expectation, while the Dipole-$k_t$ shower
(with global IF recoil) clearly has the wrong normalisation.

The reader will have noticed that in contrast with all other results
in this paper, the results here have been obtained with quite large
values of the coupling.
Furthermore the coupling has been kept fixed in the shower (and in the
associated PDF translation).
This is because it is considerably more difficult to simultaneously
explore $\as \to 0$ and $p_{tX} \to 0$ than for other
observables.
Furthermore, at large values of $\as$, had we used a running coupling,
we would have had to disentangle logarithmic effects from
power-suppressed but potentially non-negligible effects associated
with the regularisation of $\as$ near the Landau pole.

\section{Single non-global logarithms for a rapidity-slice}
\label{sec:rapidity-slice}

Many standard hadron collider observables are non-global, i.e.\
sensitive to radiation in restricted parts of angular phase space.
For example, almost any isolation criterion for leptons or photons
involves restricting the energy flow in a region around the
object.
Measurements of the top mass, as obtained from decay kinematics,
inevitably use jets that miss some radiation from the decay products.
For all of these observables, the resummation involves non-global
logarithms~\cite{Dasgupta:2001sh,Dasgupta:2002bw}, which can only be
correctly reproduced with dipole showers~\cite{Banfi:2006gy}.

To assess the ability of the shower to capture non-global logarithms (NGLs),
we compute the scalar sum of the transverse momenta $p_{ti}$ of the final-state
particles in a rapidity slice (excluding the colour-singlet
particle)~\cite{Dasgupta:2002bw,Dasgupta:2001sh}. Taking $\Delta$ to be the
half-width of a slice centred at the rapidity, $y_X$, of the
colour-singlet system, we define the observable as 
\begin{align}
  S^{\rm slice}_{\Delta} =
  \sum_{i \in \text{partons}} p_{ti} \,\Theta(\Delta-|y_i - y_X|)\,.
\end{align}
The NGLs are single-logarithmic terms of the form
$\lambda^n = \as^n L^n$, where $L = \ln(p_t/Q)$, created by the
emissions of soft large-angle gluons near the edge of the slice and we
obtain our reference resummation from the code developed for
Ref.~\cite{Caletti:2021oor}, which uses the strategy of
Ref.~\cite{Dasgupta:2001sh}.

To test the shower accuracy, we take a rapidity-slice window of
full-width 2 ($\Delta = 1$) and scan over values of
$-0.5 < \lambda=\as L < 0$.\footnote{The results are obtained with
  asymptotically small values of $\as$, so as to avoid a need for
  extrapolation.
  In the shower, collinear (initial and final-state) radiation at
  emission angles smaller than $\sim e^{-\eta_\text{max,gen}}$,
  with $\eta_\text{max,gen}=13$, is
  discarded (both from the observable calculation and from subsequent
  showering) in order
  to keep the event multiplicity under control.
  In finite-coupling runs for the PanScales showers, we have verified
  that such a cut does not impact the results.
}
Fig.~\ref{fig:rapidity-slice-a} shows PanGlobal $\betaps=0$ shower
results, as compared to the single logarithmic expectations,
illustrating perfect agreement across the full range of $\lambda$. 
Fig.~\ref{fig:rapidity-slice-b} shows results for several showers
at a fixed value of $\lambda = -0.5$, demonstrating that all showers
agree with the expected result.
The Dipole-$k_t$ showers are coloured amber because they fail to pass
fixed-order tests (see Ref.~\cite{Dasgupta:2020fwr}) and are subject
to spurious leading-colour super-leading logarithms.

\begin{figure}[t]
 \begin{subfigure}{0.5\textwidth}
 \hspace{-0.1cm}\includegraphics[height=0.335\textheight, page=5]{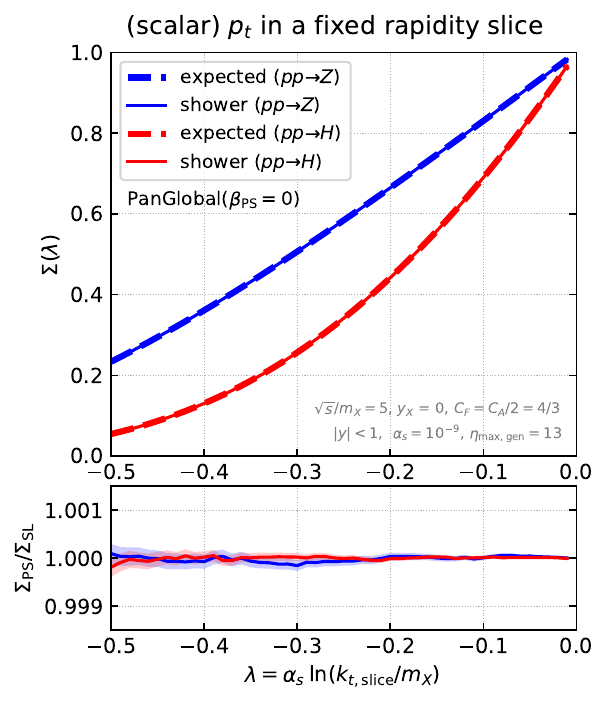}
 \caption{}
  \label{fig:rapidity-slice-a}
 \end{subfigure}%
 \begin{subfigure}{0.5\textwidth}
 \hspace{-0.7cm}\includegraphics[height=0.335\textheight, page=7]{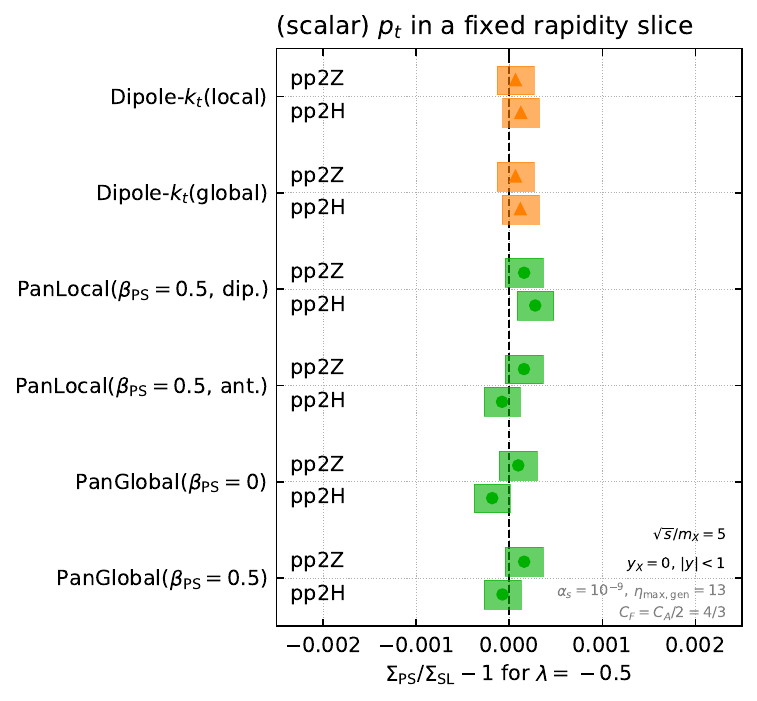}
 \caption{}
 \label{fig:rapidity-slice-b}
 \end{subfigure}
 \caption{(a) Cumulative distribution for the transverse-momentum in a
   rapidity slice of $|y-y_X| < 1$ as a 
   function of $\lambda$ for the PanGlobal $\betaps = 0$ shower. The top panel
   shows the expected (dashed) and the shower (solid) results, and the
   bottom panel shows the ratio between the two.
   (b) Difference
   between the
   shower $\Sigma_{\rm PS}$ and the expected single-logarithmic result
   ($\Sigma_{\rm SL}$) for a fixed
   value of $\lambda = -0.5$ for all showers. Colour coding is the same as
   in Fig.~\ref{fig:global-summary}. 
 }
 \label{fig:rapidity-slice}
\end{figure}

Note that this is the only one of our tests that has been performed at
leading colour ($C_F = C_A/2 = 4/3$) rather than full colour. 
The NODS scheme used elsewhere in this article is fully accurate for
non-global logarithms in colour singlet production only up to and
including $\as^2$.
Nevertheless, in $e^+e^-$ collisions it was found to be numerically
very close~\cite{Hamilton:2020rcu} to the full-colour
result~\cite{Hatta:2013iba} for non-global observables.
For a corresponding full-colour comparison in hadron collisions, one
would need an extension of the results of Ref.~\cite{Hatta:2020wre} to
include Coulomb/Glauber-gluon related (coherence-violating) $i\pi$ terms, or
of Refs.~\cite{Becher:2021zkk} or \cite{Nagy:2019bsj} to processes
without hard Born jets.

\section{Particle (or subjet) multiplicity}
\label{sec:multiplicity}

The particle multiplicity is one of the most fundamental observables
at any collider.
At hadron colliders specifically, a good understanding of the particle
multiplicity from the hard process is important in accurately
extracting the properties of the underlying event.
From a theoretical point of view, with a well-defined infrared cutoff,
the resummation structure of particle multiplicity is very similar to
that of subjet multiplicity, and our tests here effectively apply to
both.

From an analytic perspective, the resummation structure of
multiplicity differs from all other observables presented above since
its cumulative distribution cannot be written in the form of
Eq.~(\ref{eq:NLLcumulative}), and its logarithmic accuracy has to be
determined at the level of $\Sigma$ rather than $\ln \Sigma$.
The analogue of Eq.~\eqref{eq:NLLcumulative} for such
non-exponentiating observables is
\begin{align}
\label{eq:sigma-mult}
    \Sigma(L) = h_1(\as L^2) + \sqrt{\as}h_2(\as L^2) + \dots,
\end{align}
where the N$^k$DL function $\as^{k/2}h_{k+1}(\as L^2)$ resums terms of
$\as^n L^{2n-k}$. That is, the function $h_1$ captures the double
logarithmic (DL) enhancement, $h_2$ the next-to-double-logarithmic
(NDL) contribution and so on.
In the multiplicity case, the logarithm that needs to be resummed is
$L=\ln(k_{t,{\text {cut}}}/m_X)$, where, up to NDL accuracy,
$k_{t,{\text {cut}}}$ may be either a shower transverse momentum cutoff
(for particle multiplicities) or a jet algorithm transverse momentum cut
for a suitably defined subjet multiplicity.

\begin{figure}[tb]
	\centering
	\hspace{-0.75cm}\includegraphics[width=0.75\textwidth]{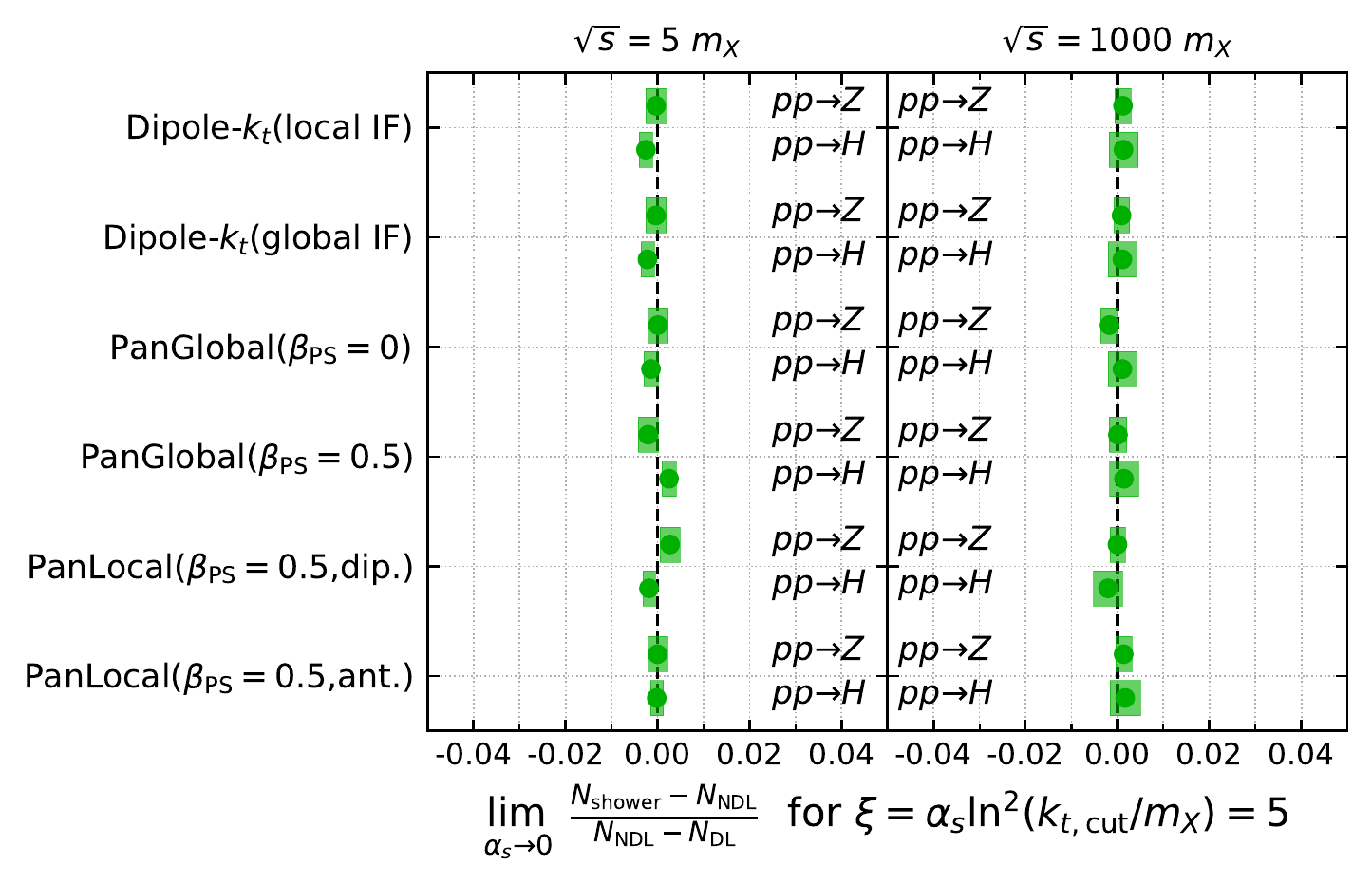}
	\caption{ Extrapolation of
		$\frac{N_\text{shower}-N_{{\rm NDL}}}{N_{\rm NDL}-N_{\rm DL}}$
		to $\alpha_s=0$ at a fixed value of $\xi=\as L^2$ for
		all showers, two different energies ($\sqrt s = 5 m_{X}$, left, and $\sqrt
		s = 1000 m_{X}$, right), and the two processes under study, i.e. $pp\to Z$
		and $pp\to H$. 
	}
	\label{fig:multiplicity-summary}
\end{figure}

Recently, the subjet multiplicity in colour singlet production has
been computed up to NDL accuracy~\cite{Medves:2022ccw} (earlier
calculations gave similar
structures~\cite{Catani:1991pm,Catani:1993yx,Forshaw:1999iv}).
In a shower context, up to NDL, it applies equally well to the number
of particles in the event ($N_\text{shower}$) when one sets the
strong coupling to zero below a given value of $k_{t,\text{cut}}$.

To test the NDL terms in Eq.~\eqref{eq:sigma-mult}, we compute the following
ratio 
\begin{equation}
\frac{N_\text{shower}-N_{{\rm NDL}}}{N_{\rm NDL}-N_{\rm DL}},
\label{eq:mult-def-ratio}
\end{equation}
which vanishes in the $\as\to 0$ limit if the shower is correct
at NDL accuracy.\footnote{Practically, we run the shower for different values
of $k_{t,\text{cut}}$, i.e.
  $\ln k_{t,\text{cut}}=\lbrace -31.25, -62.5, -125, -1000 \rbrace,$
keeping $\xi\equiv\alpha_sL^2=5$ fixed
($L = \ln k_{t,\text{cut}}/m_X$) and use all four points to perform a cubic polynomial
extrapolation down to $\alpha_s\to 0$.
The error that we quote on $N_\text{shower}$ is purely statistical.}
The result of computing Eq.~\eqref{eq:mult-def-ratio} with all showers, at two
different energies and for two different hard processes ($pp\to Z$ and $pp\to
H$) is shown in Fig.~\ref{fig:multiplicity-summary}. We observe that all showers
are consistent with the full-colour NDL expectation, within the small
statistical errors.
Relative to our other tests, the critical feature of the multiplicity
is that it probes the soft-collinear nested structure of the shower.
At NDL accuracy, it also probes the hard-collinear correction to the
splitting function, the 1-loop running of the coupling, the DGLAP
evolution of PDFs, and the colour scheme.
Since these features are common across all of our showers, no
discrepancy is expected between the PanScales showers and
Dipole-$k_t$, and none is observed.
%

\section{Exploratory phenomenological results with toy PDFs}
\label{sec:exploratory-pheno}


A proper phenomenological study of the PanScales showers would require
a number of elements that are not yet mature, such as the inclusion of
quark-mass effects and interfacing to a program such as
Pythia~\cite{Sjostrand:2014zea,Bierlich:2022pfr} so as to include
hadronisation and multi-parton interactions. 
Nevertheless, even without these effects it is still of potential
interest to examine the results from the showers in a physically
relevant regime rather than the regimes of extreme small coupling and
large logarithms used in the body of the paper.

As parton showers become more accurate, one critical element to
include in physical studies is an estimate of residual uncertainties.
In the results that follow, we will include renormalisation- and
factorisation-scale variation uncertainties, so as to provide one
measure of residual higher-order uncertainties.
However, it is important to bear in mind that these scale variations
cannot account for uncertainties associated with the showers' improper
handling of the effective matrix element in various phase-space
regions (e.g.\ the hard region, or the double-soft region). 
A study of how to do so robustly goes beyond the scope of this
section, so instead we will use the range of variation within showers
of a given logarithmic-accuracy class as an indication of such further
residual uncertainties.

Our treatment of renormalisation scale variation is inspired by
\cite{Mrenna:2016sih}, though it differs in the details.
Specifically for showers that have been established to be NLL
accurate, for an emission carrying away a momentum fraction $z$, the
emission strength is taken proportional to
\begin{equation}
  \label{eq:xmuR}
  \as(\muR^2)\left(1 + \frac{K \as(\muR^2)}{2\pi}
    + 2 \as(\muR^2) b_0 (1-z) \ln \xR \right)\,,
  \qquad
  \muR = \xR \muR^{\text{central}},
\end{equation}
where $b_0$ and $K$ are defined below  in
Appendix~\ref{app:resummation}, Eqs.~(\ref{eq:b0}).
This factor generalises the factor
$\as(\muR^2)(1 + K\as(\muR^2)/(2\pi))$ in Eq.~(2.3) of
\cite{vanBeekveld:2022zhl}, which was given for the central choice
$\muR \equiv \muR^\text{central} = \rho v {\rm
  e}^{\betaps|\bar{\eta}_Q|}$ (i.e.\ $\xR=1$ in Eq.~(\ref{eq:xmuR})),
cf.\ Eq.~(B.27) of \cite{vanBeekveld:2022zhl}.
%
%
%
The reason for including a factor $(1-z)$ in the compensation term of
Eq.~(\ref{eq:xmuR}) (i.e.\ the term proportional to $b_0$), is that it
ensures that scale compensation is active for soft emissions, $z \to 0$,
but not
for hard emissions, where one would need the higher-order ingredients
such as those from Ref.~\cite{Dasgupta:2021hbh} in order to justify
the inclusion of scale-compensating terms.
For LL showers we will include the $K$ term, but not the scale
compensation term proportional to $b_0$.
The justification for this is that a soft emission's $k_t$ is not
preserved after subsequent emissions and therefore one cannot
unambiguously identify the correct scale-dependent terms for a given
emission.
All showers use 2-loop evolution of the coupling.
We take $\as(m_Z) = 0.118$ and we implement an infrared cutoff on the
shower by setting $\as(\muR) = 0$ for $\muR < \xR \times
0.5\GeV$.

We will use a 5-flavour PDF and a 5-flavour running of the coupling,
so as to avoid complications related to the handling of flavour
thresholds. 
The PDF used for the results in this section has the initial
condition of Eq.~(\ref{eq:initial-condition}) at a scale of
$\muF = 0.5\GeV$, evolved to higher scales with HOPPET.
Aside from the issue of having $5$ flavours down to the infrared cutoff,
it is reasonably similar to a physical PDF, however not to the extent that one
can make direct comparisons with data.
Accordingly the results here should be interpreted in terms of their
broad trends, rather than specific values at any given phase space
point.
Factorisation scale variations are implemented by adding an $\ln
\xF$ term to the expression for $\ln \muF$ in Eq.~(B.1) of
\cite{vanBeekveld:2022zhl}, as used in Eq.~(2.3) of that paper.
The scale variations that we use in the plots are a 5-point set,
$(\xR, \xF) = \{ (1,1), (1/2,1), (2,1), (1,1/2), (1,2)\}$ and
we will use the envelope generated by this set as our overall scale
uncertainty band.

In the logarithmic accuracy tests in this paper, we have used the NODS
colour scheme for all showers, including the Dipole-$k_t$ showers.
Here, we retain that choice for PanScales showers, but instead use the
colour-factor from emitter scheme in the Dipole-$k_t$ case, as
this is the colour scheme adopted by standard dipole showers.

\begin{figure}
  \centering
  \includegraphics[width=0.64\textwidth]{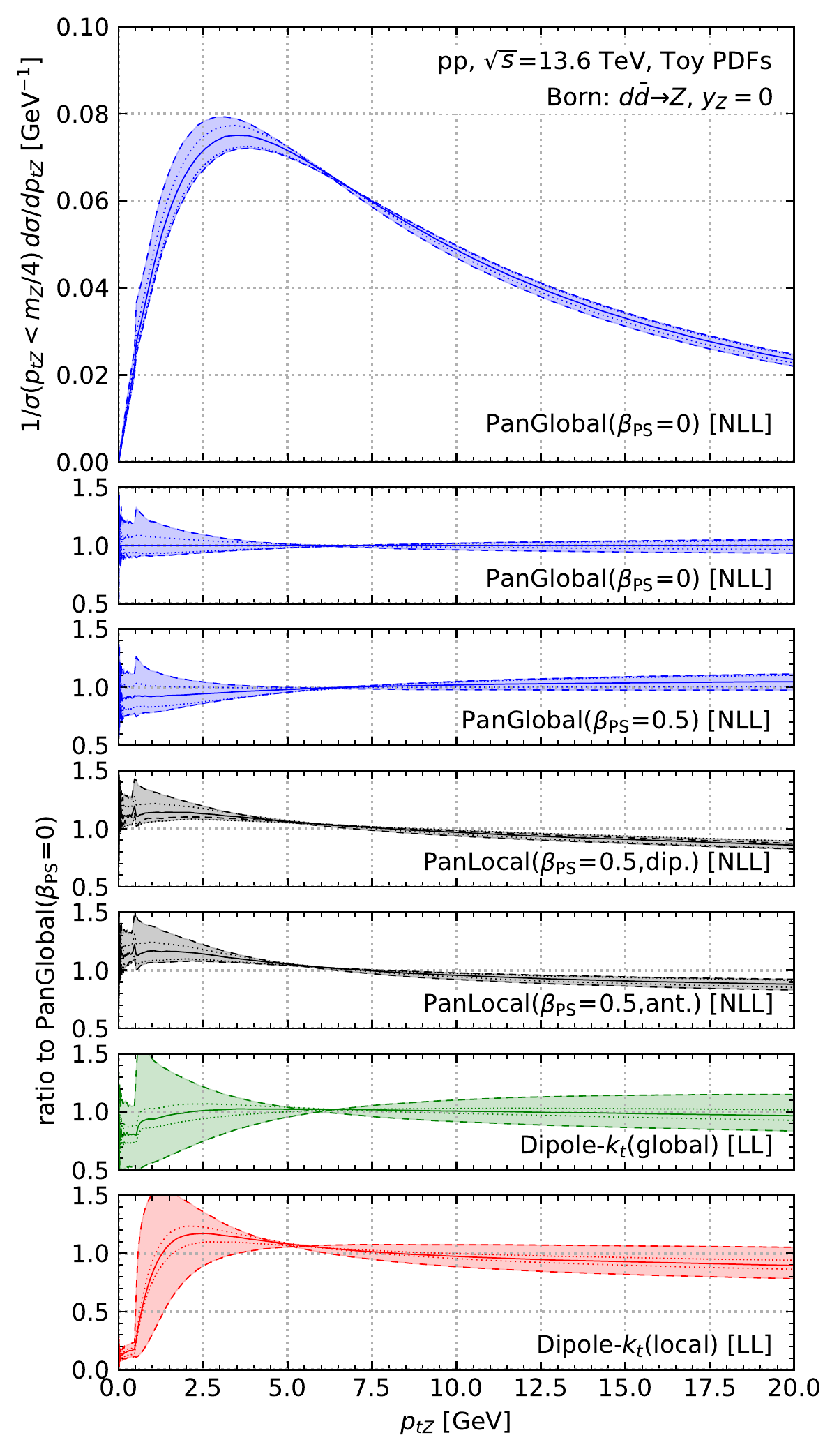}
  \caption{The $p_{tZ}$ distribution as predicted in a
    variety of parton showers.
    The plots use a semi-physical
    setup, for a $pp$ centre-of-mass energy $13.6\TeV$.
    The Born events involve $d\bar d$ scattering with a
    $Z$ rapidity of zero, and the showers use 5-flavour
    toy PDFs defined through the initial condition of
    Eq.~(\ref{eq:initial-condition}) at a scale of 0.5~GeV.
    The top panel shows the $p_{tZ}$ distribution with
    the PanGlobal ($\betaps = 0$) shower and the
    remaining panels show the ratio to that distribution
    for each of several showers.
    For each shower, the band corresponds to the envelope of the
    renormalisation  scale ($\xR$)
    variations (dashed lines)
    and factorisation scale ($\xF$) variations (dotted
    lines), as described in the 
    text.
    %
  }\label{fig:ZpTpheno}
\end{figure}

We will consider two observables: the interjet $\Delta\phi_{12}$
distribution of Section~\ref{sec:max-jet-pt} and the transverse
momentum of the $Z$-boson, $p_{tZ}$,
as discussed in Section~\ref{sec:transverse-mom}.
Let us start with the $p_{tZ}$ distribution, given its broad
phenomenological importance. 
The top panel of Fig.~\ref{fig:ZpTpheno} shows the $p_{tZ}$
distribution for the PanGlobal ($\betaps = 0$) shower, normalised to
the integral of the distribution up to $p_{tZ} = m_Z/4$.
The reason for this normalisation is to reduce sensitivity of the
results to the high $p_t$ region, where fixed-order matching would be
required to obtain a reliable prediction.
Each of the remaining panels shows the ratio of a given parton shower
(with is scale variations) to the $\xR = \xF = 1$ PanGlobal
($\betaps = 0$) result.

The first feature of Fig.~\ref{fig:ZpTpheno} that we comment on is
that the scale uncertainty bands are significantly smaller for the NLL
PanScales parton showers than for the LL dipole-$k_t$ showers.
This is because only the NLL showers include the $b_0$
scale-compensation term in the renormalisation scale uncertainty of
Eq.~(\ref{eq:xmuR}).
Next, we observe variations from one NLL shower to the next, by an
amount commensurate with the renormalisation and factorisation scale
uncertainties.
This is a consequence of different approximations for shower elements
that are beyond NLL (for example the effective treatment of the double
soft region, the specific mapping from shower scale to transverse momentum
in the hard collinear region, and the absence of matching to the hard
$Z+\text{jet}$ matrix elements).
The final comment concerns the LL showers: for the Dipole-$k_t$
(global) shower, the central value (solid curve) is rather similar to
that from the PanGlobal showers.
This is consistent with the observations in
Figs.~\ref{fig:ZHpt-distribution-a} and \ref{fig:pt-asymptotic}, from
which one expected $\sim 10\%$ agreement of
Dipole-$k_t$(global) with the PanGlobal shower, except in the deepest
part of the infrared region.
In contrast, the Dipole-$k_t$ (local) shower shows larger differences,
also as expected, notably in the different scaling behaviour at low
$p_t$ values, $p_{tZ} \lesssim 2\GeV$.
One should keep in mind that for phenomenological applications, some
of this difference might be absorbed into a tune of intrinsic
transverse momentum of partons within the proton.
However doing so might well be physically wrong, since the intrinsic
transverse momentum manifests itself in the final state through
counterbalancing transverse momentum assigned to the proton remnant
(i.e.\ concentrated just at high rapidity) rather than to soft gluon
radiation (i.e.\ spread across all rapidities).

In practical high-precision applications, parton showers results are
often reweighted so as to reproduce high-accuracy resummation and
fixed-order predictions for
$p_{tZ}$~\cite{Bizon:2019zgf,Alioli:2021qbf,Re:2021con,
  Becher:2020ugp,Camarda:2021ict,Billis:2021ecs,Ebert:2020dfc,
  Chen:2018pzu,Chen:2022cgv,Ju:2021lah,Neumann:2022lft}.
%
%
However, for a given $p_{tZ}$, such reweighting leaves the pattern of
final-state emissions unchanged.
%
%
Therefore it is also of interest to study the structure of the final
state.
We do so in Fig.~\ref{fig:delta-phi-phys}, looking at the difference
in azimuth between the two leading jets, $\Delta \phi_{12}$.
This is a close analogue of the distribution studied in
Section~\ref{sec:max-jet-pt}, but adapted so as to be
phenomenologically realistic.
Specifically, we cluster all final-state partons (excluding the
$Z$-boson) with the anti-$k_t$ algorithm~\cite{Cacciari:2008gp} with a
radius of $R=0.4$, as implemented in FastJet~3.4~\cite{Cacciari:2011ma}.
We consider only jets with $|y| < 2.5$, require at least two jets,
where the hardest has $20 < p_{t1} < 30 \GeV$ and the second hardest
has $0.3 < p_{t2}/p_{t1} < 0.5$.\footnote{This is a rather soft jet,
  and in practice one might use charged-track jets for such a study,
  so as to limit sensitivity both to pileup and to calorimeter
  fluctuations.
  One should also keep in mind that additional soft jets from
  multi-parton interactions --- not included here --- would also
  affect the results.  }
We also require a minimum rapidity separation between the jets,
$|\Delta y_{12}| > 1$, so as to reduce the impact of large-angle
$g \to gg$ (and $q\bar q$) splitting and to eliminate jet-clustering
induced artefacts associated with a suppression of the distribution
for $\Delta\phi_{12} < R$.
We then consider the distribution of $\Delta \phi_{12}$, normalised to the
number of events that passed the cuts.
This is shown in Fig.~\ref{fig:delta-phi-phys-a} for an on-shell $Z$. 
Fig.~\ref{fig:delta-phi-phys-b} shows similar results, but with two
changes: a stronger requirement on the separation between the two
leading jets, $|\Delta y_{12}| > 1.5$, to further reduce the impact of
large-angle $g \to gg/q\bar q$ splitting (uncontrolled because the
showers lack the double-soft matrix element);
and replacing on-shell $Z$-bosons with off-shell $Z$-bosons with an
invariant mass of $500 \GeV$, so that the $20\GeV$ jets are less
affected by the lack of hard matrix element corrections.

\begin{figure}
  \centering
  \begin{subfigure}{0.49\textwidth}
    \includegraphics[width=\textwidth, page=1]{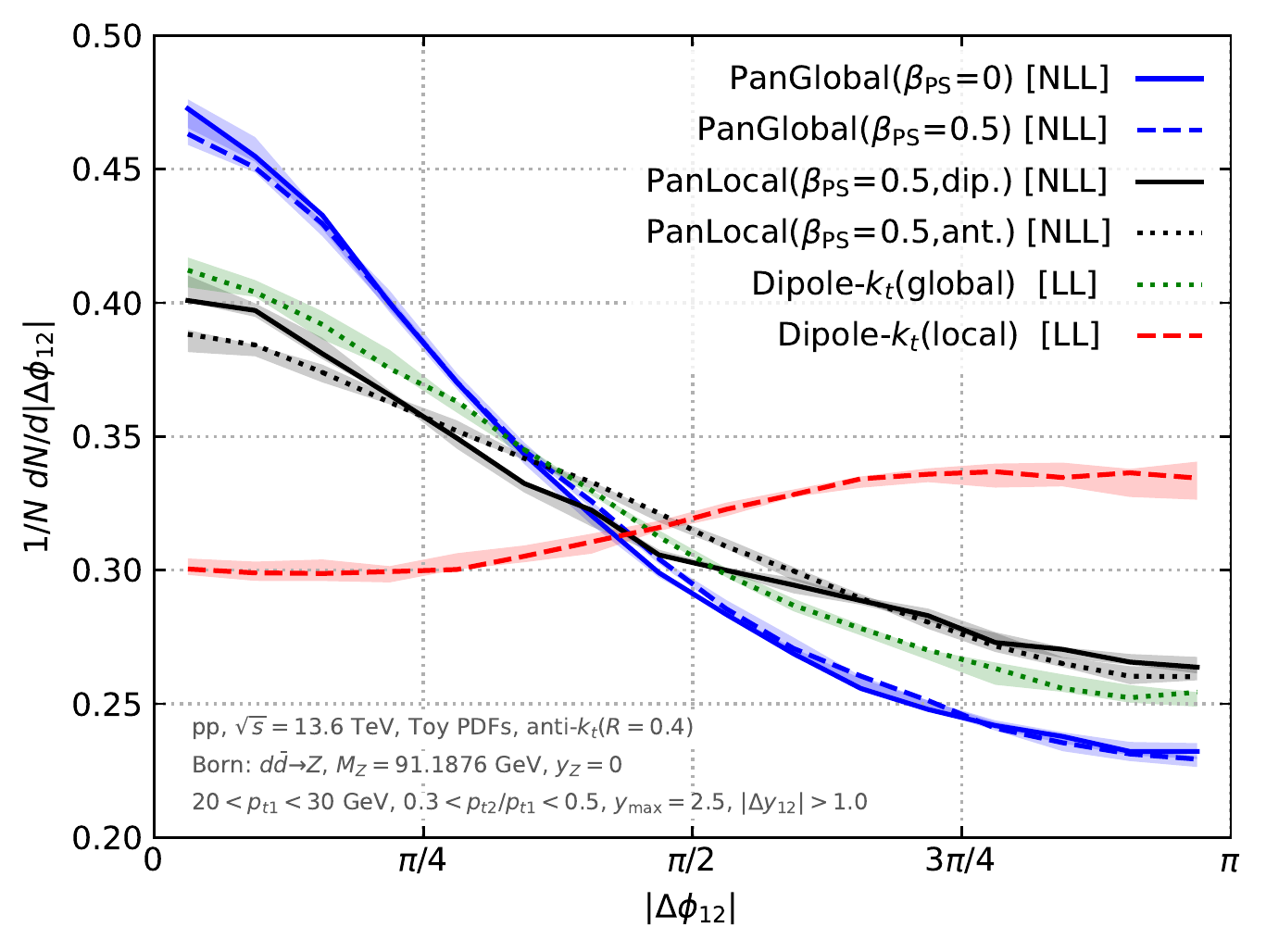}%
    \caption{\label{fig:delta-phi-phys-a}}%
  \end{subfigure}%
  \begin{subfigure}{0.49\textwidth}
    \includegraphics[width=\textwidth, page=4]{plots/pheno-delta-phi-paper.pdf}%
    \caption{\label{fig:delta-phi-phys-b}}%
  \end{subfigure}
  \caption{The $\Delta\phi_{12}$ distribution between the two leading
    jets, for events that pass the cuts described in the text, (a) for
    events on the $Z$ pole and a requirement $|\Delta y_{12}| > 1.0$,
    and (b) for events where the $Z$ is 
    off-shell, with an invariant mass of $500\GeV$, and a requirement
    $|\Delta y_{12}| > 1.5$.
    The bands correspond to the variation of renormalisation and
    factorisation scales.
    Most of the impact of this variation vanishes because the plots
    are normalised to the number of events that pass the cuts and
    instead it is the differences between showers within a given
    logarithmic accuracy class that better provides a measure of
    residual shower uncertainties.  }
  \label{fig:delta-phi-phys}
\end{figure}

The two plots in Fig.~\ref{fig:delta-phi-phys} can be compared to
their analogue in the asymptotic logarithmic limit,
Fig.~\ref{fig:maxpt-deltaphi-b}.
One caveat is that the former normalises to the cross section for the
jets to pass the transverse momentum and rapidity selection cuts (as
would most likely be done experimentally), while the latter normalises
to the asymptotic NLL expectation of Eq.~(\ref{eq:deltaphinll}), which
is simple only in the absence of rapidity cuts on the jets.
A first feature to comment on is that in
Fig.~\ref{fig:delta-phi-phys}, the PanScales showers are not flat in
$\Delta\phi_{12}$, unlike the case in Fig.~\ref{fig:maxpt-deltaphi-b}.
This is because in a non-negligible fraction of the events with two
jets passing the cuts, those two jets effectively came from a
large-angle $g \to gg$ (or $q\bar q$) splitting, and so have
$\Delta\phi_{12}$ close to $0$, resulting in the enhancement seen in
that region.
We have verified that increasing the $\Delta y_{12}$ cut, e.g.\ to
$2$, leads to a degree of flattening of the distribution for all of
the PanScales showers, as does eliminating the $|y| < 2.5$ selection
on the jets (which increases the relative contribution of
configurations with large rapidity separations).%
\footnote{
  For comparison, the $\as \to 0$ limit used in the NLL tests of
  Fig.~\ref{fig:maxpt-deltaphi-b} effectively ensures that the two
  jets are nearly always well separated in rapidity.  }
\logbook{0c03bbe3f0}{To explore various options, use explorer.html in
  2020-eeshower/analyses/pp-analyses/results/pheno/html-explorer
  (see the README to set up a temporary local http server).
  For a set of static plots, see
  2020-eeshower/analyses/pp-analyses/results/pheno/pheno-delta-phi.pdf.
  For Pythia studies with/without MPI, see
  2020-eeshower/analyses/pp-analyses/pythia/. 
}%
Among the PanScales showers there is some spread between the showers
in Fig.~\ref{fig:delta-phi-phys-a}, notably between the PanGlobal
variants on one hand and the PanLocal variants on the other.
This spread largely vanishes when probing a more asymptotic kinematic
region, Fig.~\ref{fig:delta-phi-phys-b}, and further investigation
shows that the reduction of spread stems both from the increase in
colour singlet mass and the $\Delta y_{12}$ requirement.
Note that the renormalisation and factorisation scale variation are far from
encompassing the spread in Fig.~\ref{fig:delta-phi-phys-a} (in part
because the scale variations are divided out by the normalisation).
This is a sign that scale variation alone is not sufficient for
probing the uncertainties in parton showers, as in many other
contexts, and that one also needs to investigate uncertainties
related to uncontrolled limits of matrix elements.\footnote{This point
  was touched on in Ref.~\cite{Mrenna:2016sih}, but should, we
  believe, be further explored.}

To close our discussion, we turn to the Dipole-$k_t$ results in
Fig.~\ref{fig:delta-phi-phys}. 
The variant with local-IF recoil has a substantially different shape
from the NLL showers.
Even though the shape differs from the asymptotic limit in
Fig.~\ref{fig:maxpt-deltaphi-b} (again because of residual
$g \to gg/q\bar q$ splittings), the enhancement at
$\Delta\phi_{12}\simeq \pi$, relative to the NLL showers, is
qualitatively as expected from that plot.
In contrast, we see that the Dipole-$k_t$ variant with global-IF
recoil in Fig.~\ref{fig:delta-phi-phys-a} is fairly similar to the NLL
showers.
In this kinematic region, the logarithms are not
yet very large.
As a result the smaller LL versus NLL differences (for this observable)
of the Dipole-$k_t$(global) shower as compared to Dipole-$k_t$(local),
are commensurate with the beyond-NLL differences between PanScales
showers.
However, the results in Fig.~\ref{fig:delta-phi-phys-a}, if taken
alone, would give a false sense of confidence in the phenomenological
adequacy of the Dipole-$k_t$(global) shower for this observable.
In particular exploring a more asymptotic kinematic region, as in
Fig.~\ref{fig:delta-phi-phys-b}, reveals clear differences also
between Dipole-$k_t$(global) and the NLL PanScales showers.

\section{Conclusions}
\label{sec:conclusions}

In this article, we have carried out over a dozen distinct all-order
tests of the logarithmic accuracy of parton showers for colour-singlet
production at hadron colliders.

On one hand, these tests were designed to probe distinct classes of
next-to-leading logarithmic effects, covering all of the main aspects
that a shower should be able to handle.
On the other, each of the observables also connects with important
phenomenological aspects of LHC physics.
The tests probed nested emissions in the hard collinear region (DGLAP
tests of Section~\ref{sec:pdf-origin});\footnote{With the exception of
  spin-correlation tests, for which we are not aware of any all-order
  results, besides those that could be obtained with our code.}
nested emissions in the soft large-angle region (non-global
observables of Section~\ref{sec:rapidity-slice});
nested emissions in
both the soft and collinear regions (multiplicities of
Section~\ref{sec:multiplicity});
and the higher-order structure of double logarithmic Sudakov
resummation, including both recoil and the scale and scheme of the
coupling in the Sudakov form factor (global observables of
Sections~\ref{sec:global-obs} and \ref{sec:transverse-mom}).
All of these tests were carried out with the NODS colour scheme of
Ref.~\cite{Hamilton:2020rcu}, and with comparisons to full-colour
resummation (with the exception of non-global observables).

For the PanLocal shower (with $\betaps = 0.5$) and the PanGlobal
showers ($\betaps=0$ and $0.5$), all tests were successful.
For the Dipole-$k_t$ showers, we considered two variants, one with
dipole-local (``local''), the other with event-wide (``global'')
recoil in initial--final dipoles.
Both had visible discrepancies relative to NLL for all global
observables that connect directly with transverse momentum
measurements.
This includes the jet veto acceptance (Fig.~\ref{fig:maxpt-deltaphi-a}),
a number of generic global observables (Fig.~\ref{fig:global-summary})
and the colour-singlet transverse momentum distribution
(Figs.~\ref{fig:ZHpt-distribution} and \ref{fig:ptZ-scaling}).
Note that our Dipole-$k_t$ tests used our NODS colour scheme.
Had we used the colour treatment that is effectively standard for
dipole showers (colour-factor-from-emitter in the language of
Ref.~\cite{Hamilton:2020rcu}), we would also have seen subleading-$\nc$
issues in the Dipole-$k_t$ showers at LL for $\beta_\text{obs} >0$
global observables, NLL for $\beta_\text{obs}=0$ and DL for
multiplicities. 

A number of steps remain for practical phenomenological
applications of the PanScales showers.
These include the matching to fixed-order calculations, the extension
of our validations and tests to hadron-collider processes with
final-state jets, the inclusion of finite quark masses and the
interface to hadronisation and multi-particle-interaction models.
Nevertheless, the advances presented here provide an important step in
the formulation and validation of NLL-accurate showers for hadron
collisions.
The first exploration of the phenomenological impact of our NLL
showers in Section~\ref{sec:exploratory-pheno} shows some of the
potential benefits from the control of logarithmic accuracy.


\section*{Acknowledgements}

We are grateful to our PanScales collaborators (%
Mrinal Dasgupta,
Fr\'ed\'eric Dreyer,
Basem El-Menoufi,
Jack Helliwell,
Alexander Karlberg,
Rok Medves,
Pier Monni,
Ludovic Scyboz,
Scarlett Woolnough%
),
for their work on the code, the underlying
philosophy of the approach and comments on this manuscript.

This work was supported
by a Royal Society Research Professorship
(RP$\backslash$R1$\backslash$180112) (MvB, GPS),
by the European Research Council (ERC) under the European Union’s
Horizon 2020 research and innovation programme (grant agreement No.\
788223, PanScales) (SFR, KH, GPS, GS, ASO, RV), 
and by the Science and Technology Facilities Council (STFC) under
grants ST/T000856/1 (KH) and ST/T000864/1 (MvB, GPS).

\appendix
\section{Parton distribution functions}
\label{sec:choice-pdfs} 
The inclusion of a ratio of parton distribution functions in the
branching kernel for initial-state emissions is a vital component of a
complete hadronic parton shower.
While the PanScales showers follow the backwards evolution in much the
same way that all other widely-used showers do, the numerical demands
on the implementation are often of a very different order.
Below, we discuss some of the
numerical details in our handling of PDFs.
Appendix~\ref{sec:pdf-ratio-overestimate} outlines our procedure for
overestimating the PDF ratio that appears in the branching kernels.
Appendix~\ref{sec:pdf-scale} outlines how we obtain PDFs for use in
limits with extreme values of the logarithm and tiny values of $\as$.
Finally, Appendix~\ref{app:pdfchoice} provides the specific functional
form that we use for our PDFs when testing logarithmic accuracy.

\subsection{Overestimating the PDF ratio}
\label{sec:pdf-ratio-overestimate}

In the PanScales formalism, the differential branching probability for
initial-state emissions may be written as 
\begin{multline}
    \label{eq:dipole-prob}
        {\rm d} \mathcal{P}_{\itilde \jtilde \to ijk} =
        \frac{\as(k_{\perp}^2)}{2 \pi}
        \left( 1 + \frac{\as(k_{\perp}^2) K}{2 \pi} \right)
        \frac{{\rm d} v^2}{v^2} {\rm d}
        \eta \frac{{\rm d}\varphi}{2\pi} \times \\
         \times \frac{x_i f_i(x_i, \mu^2)}{\tilde{x}_i f_{\itilde}(\tilde{x}_i,\mu^2)} \frac{x_j f_j(x_j, \mu^2)}{\tilde{x}_j f_{\jtilde}(\tilde{x}_j,\mu^2)  } \left[ g(\eta) z_i P^{\rm {IS/FS}}_{ik} (z_i) + g(-\eta) z_j P^{\rm {IS/FS}}_{jk} (z_j) \right] \,,
\end{multline}
where $\mu$ is the factorisation scale to be used in the PDFs and the
rest of the notation is as in Ref.~\cite{vanBeekveld:2022zhl}.
To implement these branchings in the shower using the standard veto
algorithm, an overestimate is required for the branching probability.
In the absence of the PDF
ratios (i.e.\ for final-state branchings), the branching probability is
easily overestimated by a constant, $\as C_A/\pi$.
This is not quite as
straightforward anymore when the PDF ratio is included. The shower is
maximally efficient if the overestimate is as tight as possible, but
it will not produce the correct distributions if regions exist where
the branching probability is not correctly overestimated.

We implement a solution that maintains the simplicity of the final-state case as
much as possible by introducing a further overhead factor $C^{\text{PDF}}(\tilde{x},
\itilde)$ that depends on the current longitudinal momentum fraction
$\tilde{x}$ of an initial-state parton, as well as its flavour $\itilde$.
The generic overestimate constant $\as C_A/\pi$ is multiplied by this
overhead factor, and the acceptance probability is divided by it.

The overhead factor is evaluated by filling grids with values of
$C^{\text{PDF}}(\tilde{x}, \itilde)$ for all $\itilde$s and for
equally-spaced values of $-\ln 2\tilde{x}$ for $\tilde{x} < 1/2$, or $-\ln
2(1-\tilde{x})$ for $\tilde{x} > 1/2$. Then, for every pair $(\tilde{x},
\itilde)$ in the grid, a secondary grid scan is performed over $x >
\tilde{x}$ and all factorisation scales that can be accessed in the
shower, so as to identify the maximum
branching weight given in Eq.~\eqref{eq:dipole-prob}.
Once a maximum is found, another grid search is performed in the cell
of the previously-identified maximum. This process is repeated four
times and the result
is then multiplied by a margin factor $1.2$ and stored in the
$C^{\text{PDF}}(\tilde{x}, \itilde)$ grid.

At the beginning of the shower evolution and after every shower
branching, an adequate overhead factor can then be determined by
probing the $C^{\text{PDF}}(\tilde{x}, \itilde)$-grid at the current
longitudinal momentum fraction and flavour of the initial-state
partons. While this procedure is not guaranteed to determine an
overestimate over all of phase space, we find that the modest margin
factor of $1.2$ avoids any issues without significant detriment to
efficiency. This method is applicable in principle to any reasonably
well-behaved PDF set, as long as we remain in a region of fixed number
of flavours, i.e.\ stay away from potential mass
threshold effects that can cause the overhead factor to
diverge.\footnote{
  %
  To understand the nature of the difficulty around heavy-flavour
  thresholds, we imagine using an NLO 
  PDF, such that the heavy-quark distribution is zero below the
  heavy-quark mass $m_Q$ and starts evolving from scale $\mu = m_Q$.
  As a result the heavy-quark PDF scales as
  $\ln \mu/m_Q \simeq (\mu - m_Q)/m_Q$ in the immediate vicinity above
  $m_Q$.
  This is problematic, because for a heavy-quark ($\itilde$) that
  backwards evolves to a gluon ($i$), as has to happen if the
  evolution scale is close to the heavy-quark threshold, the PDF ratio
  in Eq.~(\ref{eq:dipole-prob}) diverges as $1/(\mu - m_Q)$.
  This cannot be compensated for by a tabulated overhead factor.
  We illustrate  the type of solution that we might consider with the
  example of a transverse-momentum ordered shower where 
  $v \equiv \kappa_\perp \equiv \mu$.
  For a dipole containing an initial-state heavy quark, we could
  perform a change of variable, replacing our current logarithmic
  generation variable $\ln \kappa_\perp \equiv \ln \mu$, with generation of
  $\ln (\mu - m_Q)$, which would entail the inclusion of a
  Jacobian factor $\mu/m_Q - 1$.
  That Jacobian would then cancel the $1/(\mu - m_Q)$ factor that
  arises from the PDF ratio in Eq.~(\ref{eq:dipole-prob}).
  Another possibility would be that for heavy-quark PDFs, we replace
  the $\mu^2 \to \mu^2 + m_Q^2$ in the scale of the PDF in
  Eq.~(\ref{eq:dipole-prob}) and tabulate the (large, but finite)
  overhead factor all the way down the shower cutoff.
  The first scheme would ensure that a heavy quark always branches
  back to a gluon by the time the shower crosses $\mu = m_Q$,
  while the second scheme would allow for intrinsic heavy flavour at
  the shower cutoff scale. 
  %
  We have not implemented either of these solutions as yet, and so
  defer further discussion both of their practicalities and
  phenomenological behaviour to future work. 
}
For our all-order tests, we make use of the toy PDFs given in
Appendix~\ref{app:pdfchoice}. 

\subsection{PDFs at extreme scales}
\label{sec:pdf-scale}
Standard PDF evolution tools are not well-suited to our requirements
of being able to evaluate PDF ratios in the limit where $\as \to 0$
with $\as L$ fixed.
At the accuracy we intend to probe, NLL or NDL, it is sufficient to
make use of PDFs with purely collinear, single-logarithmic DGLAP
evolution.
This means that only leading-order splitting functions and 1-loop
running of $\as$ are required.\footnote{The shower itself still needs
  a $2$-loop running coupling.
  This is critical for NLL accuracy in the soft-collinear region, a
  region that does not significantly contribute to PDF evolution.  }
In this situation, the PDF evolution is a function purely of an
evolution time parameter $t = \as L$.
We can leverage this fact to evaluate PDFs at extremely small scales
without having to explicitly perform the DGLAP evolution to those
scales.

In what follows we use $\mu$ to denote a factorisation scale within
the shower (which operates over asymptotic scales) and $\mu_\pdf$ to
denote a factorisation scale in the PDF evolution (which operates
over standard physical scales).
We start by generating a five-flavour, one-loop PDF set between a
lower scale $\mu_{\pdf,0}$ and a high scale $\mu_{\pdf,1}$, with a value of
$\aspdf(\mu_\pdf^2)$ similar to the physical value.
This task can be handled by DGLAP evolution codes. We choose to use
the HOPPET library \cite{Salam:2008qg}.
The PDF scale $\mu_{\pdf,1}$ is then mapped onto the shower initial hard
transverse momentum scale $\mu_{1}$.
Lower scales are then related by
\begin{equation} \label{eq:evolution-time}
    t_\pdf(\mu_\pdf|\mu_{\pdf,1}) = t_\text{shower}(\mu|\mu_1)\,.
\end{equation}
At the NLL accuracy that we are aiming for, where it is sufficient to use 
one-loop DGLAP evolution, the left-hand side is given by 
\begin{equation} \label{eq:evolution-time-one-loop}
    t_\pdf(\mu_\pdf|\mu_{\pdf,1}) = \int^{\mu_{\pdf,1}}_{\mu_\pdf} \frac{{\rm d}q}{q} \frac{\aspdf(q^{2})}{\pi} = \frac{-\ln\left(1 + \beta_0 \, \aspdf(\mu_{\pdf,1}^{2}) \ln \mu_\pdf^2/\mu_{\pdf,1}^2\right)}{2 \pi \beta_0}\,,
\end{equation}
and we have
\begin{equation}
    \mu_\pdf(\mu) = \mu_{\pdf,1} \exp\left(\frac{\as(\mu_1^{2})}{\aspdf(\mu_{\pdf,1}^{2})} \ln \frac{\mu}{\mu_{1}}\right),
\end{equation}
where $\as(\mu_{1}^2)$ is the value of the coupling used in the shower
evolution at the hard scale $(\mu_{1})$. The shower PDF can then be
evaluated as
\begin{equation}
    f_i(x, \mu^2_\pdf(\mu))\,.
\end{equation}
The choice of the numerical values of $\mu_{\pdf,1}$, $\mu_{\pdf,0}$
and $\aspdf(\mu_{\pdf,1}^{2})$ is somewhat arbitrary, only requiring that
the DGLAP evolution is performed over a range that is wide enough to
cover the kinematic range of the shower, and that the evolution is
numerically stable.
The above procedure then facilitates consistent comparison of shower
runs with a variety of values of $\as(\mu_1^{2})$, as long as the
upper boundary of the shower, $\mu_{1}$, always remains anchored
to the same PDF scale $\mu_{\pdf,1}$.

\subsection{PDF choice}
\label{app:pdfchoice}
We employ a toy PDF set whose functional form is defined at the
starting scale for the evolution $\mu_{\pdf,0} = 1\GeV$, with the
coupling at that scale set to $\as(\mu_{\pdf,0}^2) = 0.5$.
For the gluon PDF at that scale we take
\begin{subequations}
  \label{eq:initial-condition}
  \begin{align}
    g(x) &= N_g \,x^{\beta}\,(1-x)^5\,,
  \end{align}
  with $\beta= -0.1$ and $N_g = 1.7$. For the quark PDFs we define
  \begin{align}
    u_v(x) &= N_{u_v}\,x^{\alpha}\,(1-x)^3\,, \\
    d_v(x) &= N_{d_v}\,x^{\alpha}\,(1-x)^4\,, \\
    \tilde{d}(x) &= N_{\bar{d}}\,x^{\beta}\,(1-x)^6\,, \\
    \tilde{u}(x) &= N_{\bar{d}}\,x^{\beta}\,(1-x)^7\,,
  \end{align}
  with $\alpha = 0.8$, $N_{u_v} = 5.1072$, $N_{d_v} = 3.06432$ and
  $N_{\bar{d}} = 0.1939875$. We then use
  \begin{align}
    u(x) &= u_v(x) + 0.8\tilde{u}(x)\,, \\
    d(x) &= d_v(x) + 0.8\tilde{d}(x)\,, \\
    \bar{u}(x) &= 0.8\tilde{u}(x)\,, \\
    \bar{d}(x) &= 0.8\tilde{d}(x)\,, \\
    s(x) &= \bar{s}(x) = 0.2(d(x) + \tilde{d}(x))\,, \\
    c(x) &= \bar{c}(x) = b(x) = \bar{b}(x) = 0.15(\tilde{u}(x) + \tilde{d}(x))\,,\\
    t(x) &= \bar{t}(x) = 0\,.
  \end{align}
\end{subequations}
In the above equations the forms for the PDFs are all implicitly to be
understood as being at the factorisation scale $\mu_{\pdf,0}$.
The PDF uses $n_f = 5$ light flavours, as with the rest of our
results in this paper.
For the purposes of mapping shower scales to PDF scales, as in
Appendix~\ref{sec:pdf-scale}, we use $\mu_{\pdf,1} = 10^7\GeV$.%
\footnote{%
  For Fig.~\ref{fig:ptZ-scaling}, because of the use of fixed coupling
  in the shower, we needed a particularly large range of PDF evolution
  (which uses 1-loop running of the coupling), and used
  $\mu_{\pdf,0} = 0.5\GeV$, $\as(\mu_{\pdf,0}^2)=1.2$,
  $\mu_{\pdf,1}=10^{20}\GeV$.
  An alternative solution would have been to adapt HOPPET to have the
  option of evolving the PDFs with a fixed coupling.  }

\section{Resummation formulae}
\label{app:resummation}

In this appendix we summarise the NLL analytic resummation expressions 
used in Section~\ref{sec:global-obs}.\footnote{Here we do not discuss
  the question of coherence-violating (``super-leading'')
  logarithms~\cite{Forshaw:2006fk,Catani:2011st}, whose role in
  resummations for colour-singlet production processes at hadron colliders
  remains to be further investigated
  (see also footnote~15 of Ref.\cite{vanBeekveld:2022zhl}).
}
We consider a continuously global
observable~\cite{Banfi:2004yd} that, for a single soft or collinear emission
with transverse momentum $k_t$ and rapidity $y$ takes the form
\begin{equation}
O = \frac{k_t}{Q} e^{-\beta_{\text{obs}}|y - y_X|},
\end{equation}
with $0 \leq \beta_{\text{obs}} \leq 1$ and $y_X$ the rapidity of the
massive colour-singlet boson.
The probability that the observable is smaller than $e^{L}$, where $L$ is taken to be large and negative,
can be written at NLL accuracy as
\begin{equation}
\label{eq:master-nkll}
 \Sigma(\as, \as L) = \exp\left[-L g_1(\as L) + g_2(\as L) +\mathcal{O}(\as^n L^{n-1})\right] 
 \end{equation}
The  $g_1$-function contains the LL terms and reads
\begin{subequations}
\begin{align}
    g^{\beta_\text{obs} = 0}_1&= 2 C_i \left[\frac{1}{2 \pi b_0\bar\lambda}\left(2 \bar\lambda + \ln(1-2\bar\lambda)\right)\right]\,,
    \label{eq:g1-b0} \\
    g^{\beta_\text{obs} \neq 0}_1 &=  2 C_i \left[\frac{1}{2\pi b_0\bar\lambda\beta_\text{obs}}\left((1 + \beta_\text{obs} - 2 \bar\lambda) \ln\left(1-\frac{2\bar\lambda}{1+\beta_\text{obs}}\right) - (1-2 \bar\lambda)\ln(1-2\bar\lambda)\right)\right] 
    \label{eq:g1-bn0}\,,
\end{align}
\end{subequations}
with 
\begin{align}
  \label{eq:b0}
b_0 = \frac{11 C_A - 4 n_f T_R}{12 \pi}\,, \quad\quad \bar\lambda = - b_0 \as L = -b_0 \lambda\,,
\end{align}
and $C_i = C_A (C_F)$ for Higgs ($Z$) production.

The NLL corrections in Eq.~\eqref{eq:master-nkll} are resummed in the $g_2$-function.
This function contains contributions from (i) soft-collinear emissions $r_2$,
(ii) hard-collinear emissions $T$, (iii) the PDF evolution
$\mathcal{L}$, and (iv) a factor
$\mathcal{F}_{\beta_{\rm obs}}$ that accounts for the way the
observable depends on multiple emissions.
It takes the general form
\begin{eqnarray}
    \label{eq:g2-nll}
g_2 = -2C_i r_2(\bar\lambda) - 2 C_i B_i T\left(\bar\lambda\right) +
\ln \mathcal{L} + \ln \mathcal{F}_{\beta_\text{obs}}\,,
\end{eqnarray}
with $C_i$ the same as above, $B_i = B_q = -3/4$ for quarks and $B_g = (-11 C_A
+ 4 n_f T_R)/ 12 C_A$ for gluons. The NLL contribution from soft-collinear
emissions reads
\begin{subequations}
\begin{align}
r_2^{\beta_\text{obs} = 0}&= \frac{1}{2\pi b_0^2}\left[\frac{K}{2\pi}\left(\ln(1 - 2\bar\lambda) + \frac{2\bar\lambda}{1-2\bar\lambda}\right) - \frac{b_1}{b_0}\left(\frac{1}{2}\ln^2(1-2\bar\lambda) + \frac{\ln(1-2\bar\lambda) + 2\bar\lambda}{1-2\bar\lambda}\right)\right]\,, \\
r_2^{\beta_\text{obs} \neq 0} &= \frac{1}{2\pi b_0^2 \beta_\text{obs}}\Bigg[\frac{K}{2\pi}\left((1+\beta_\text{obs})\ln\left(1 - \frac{2\bar\lambda}{1+\beta_\text{obs}}\right)- \ln(1 - 2\bar\lambda)\right)\nonumber  \\
&+ \frac{b_1}{b_0}\Bigg(\frac{1}{2}\ln^2(1-2\bar\lambda) - \frac{1}{2}(1+\beta_\text{obs})\ln^2\left(1 - \frac{2\bar\lambda}{1+\beta_\text{obs}}\right) + \ln(1-2\bar\lambda)  \nonumber \\
& - (1 + \beta_\text{obs} )\ln\left(1 - \frac{2\bar\lambda}{1+\beta_\text{obs}}\right)  \Bigg)\Bigg] ,
\end{align}
\end{subequations}
with
\begin{align}
b_1 =    \frac{17 C_A^2 - 10 C_A n_f T_R - 6 C_F n_f T_R}{24\pi^2}\,, \qquad 
K =  \left(\frac{67}{18}-\frac{\pi^2}{6}\right) C_A-\frac{10}{9} n_f T_R\, ,
\end{align}
while the corresponding term for hard-collinear emissions is given by
\begin{align}
T= -\frac{1}{\pi b_0}\ln\left(1-\frac{2 \bar\lambda}{1+\beta_{\rm obs}}\right)\,.
\end{align} 
The contribution arising from the PDFs evolution in processes with two coloured legs in the initial state
($\ell_{1,2}$) is given by 
\begin{align}
  \label{eq:lumi-factor}
\ln \mathcal{L} = \ln\left(\frac{f_{\ell_1}\!\left(x_1,  Q^{2}
  e^{2L/(1+\beta_{\text{obs}})}\right)}{f_{\ell_1}(x_1,Q^{2})}\right)
  + \ln\left(\frac{f_{\ell_2}\!\left(x_2, Q^{2} e^{2L/(1+\beta_{\text{obs}})}\right)}{f_{\ell_2}(x_2,Q^{2})}\right),
\end{align}
and
$f_{\ell_i}(x_i, \mu^2)$ the PDF for flavour $\ell_i$ evaluated for a light cone momentum fraction $x_i$ at the factorisation scale
$\mu$.
In conjunction with our PDF mapping from Appendix~\ref{sec:pdf-scale},
Eq.~(\ref{eq:lumi-factor}) depends on the logarithm of the observable
only through the value of $\lambda$.
The last term of Eq.~\eqref{eq:g2-nll} depends on the type of observable. For an
additive observable we have
\begin{align}
\label{eq:f-sum-obs}
\ln \mathcal{F}_{\beta_\text{obs}}^{S} = -\gamma_E R'(\bar\lambda) - \ln\Gamma\left(1 + R'(\bar\lambda)\right)\,,
\end{align}
with $R'(\bar\lambda)$ defined as $\partial_L\left(-L g_1(\bar\lambda)\right)$, i.e.
\begin{subequations}
\begin{align}
    R'_{\beta_\text{obs} = 0}(\bar\lambda) &= \frac{4C_i}{\pi b_0}
                                    \frac{\bar\lambda}{1-2\bar\lambda}\,,
                                    \label{eq:Rp-beta0}
  \\
    R'_{\beta_\text{obs} \neq 0}(\bar\lambda) &= \frac{2C_i}{\pi b_0\beta_\text{obs}} \left[\ln\left(1-\frac{2\bar\lambda}{1+\beta_\text{obs}}\right) - \ln(1-2\bar\lambda) \right].
\end{align}
\end{subequations}
For observables involving a maximum among jets, i.e.\ the $M_{j,
  \beta_\text{obs}}$ of Eq.~(\ref{eq:global-obs-Mj}), we have 
\begin{equation}
\ln \mathcal{F}_{\beta_\text{obs}}^{M} = 0\,.
\end{equation}
The transverse momentum of the colour singlet also belongs to the class of
global observables with $\beta_\text{obs}=0$. In this case, the observable-dependent correction reads
\begin{align}
  \ln \mathcal{F}_{\beta_\text{obs}}^{p_{tX}} = - \gamma_E R'(\bar\lambda) - \ln\Gamma\left(1 + R'(\bar\lambda)/2\right) + \ln\Gamma\left(1 - R'(\bar\lambda)/2\right)\,,
\label{eq:curlyF_Vpt}
\end{align}
which has pole at $R'(\bar\lambda)=2$.


\bibliographystyle{JHEP}
\bibliography{MC}

\end{document}
